\documentclass[12pt]{article}

\usepackage{tikz}
\usetikzlibrary{matrix,arrows,decorations.pathmorphing,decorations.markings,patterns}

\usepackage{jheppub}

\usepackage{amsmath,amssymb,euscript,array,mathrsfs,amsfonts}
\usepackage{epsfig}

\newcommand{\field}[1]{\mathbb{#1}} 
\newcommand*{\Dsl}[0]{{\rlap{\kern2.25pt /}{D}}}
\newcommand*{\Asl}[0]{{\rlap{\kern2.25pt /}{A}}}
\newcommand*{\dsl}[0]{{\rlap{\kern0.5pt /}{\partial}}}
\newcommand*{\xisl}[0]{{\rlap{\kern0.5pt /}{\xi}}}
\newcommand*{\asl}[0]{{\rlap{\kern0.5pt /}{a}}}
\newcommand*{\bsl}[0]{{\rlap{\kern0.5pt /}{b}}}

\newcommand{\lapprox}{\raisebox{-0.5ex}{$\
\stackrel{\textstyle<}{\textstyle\sim}\ $}}
\newcommand{\gapprox}{\raisebox{-0.5ex}{$\
\stackrel{\textstyle>}{\textstyle\sim}\ $}}

\def\Dslash{\,\,{\raise.15ex\hbox{/}\mkern-12mu D}}

\newcommand{\SP}[1]{\begin{equation}\begin{split} #1
\end{split}\end{equation}}

\newcommand{\Tr}{\operatorname{Tr}}

\def\B0{{\boldsymbol 0}}

\def\Tr{{\rm Tr}}

\def\Dbarslash{\,\,{\raise.15ex\hbox{/}\mkern-12mu {\bar D}}}
\def\Dslash{\,\,{\raise.15ex\hbox{/}\mkern-12mu D}}
\def\delslash{\,\,{\raise.15ex\hbox{/}\mkern-9mu \partial}}
\def\delbarslash{\,\,{\raise.15ex\hbox{/}\mkern-9mu {\bar\partial}}}

\newcommand{\EQ}[1]{\begin{equation}\begin{split} #1
\end{split}\end{equation}}

\title{Deconfinement transitions of large N QCD with chemical potential at weak and strong coupling}
\author[a]{Timothy J. Hollowood}
\author[b]{and Joyce C. Myers}
\affiliation[a]{Department of Physics, Swansea University, Singleton Park, Swansea SA2 8PP, U.K.}
\affiliation[b]{University of Groningen,
  Centre for Theoretical Physics, 9747 AG, Groningen, The Netherlands}
\emailAdd{t.hollowood@swansea.ac.uk}
\emailAdd{j.c.myers@rug.nl}
\abstract{We calculate the deconfinement line of transitions for large $N_c$ QCD at finite temperature and chemical potential in two different regimes: weak coupling in the continuum, and, strong coupling on the lattice, working in the limit where $N_f$ is of order $N_c$. In the first regime we extend previous weak-coupling results from one-loop perturbation theory on $S^1 \times S^3$ to higher temperatures, where the theory reduces to a matrix model, analogous to that of Gross, Witten, and Wadia. We obtain the line of transitions that extends from the temperature-axis, where to a first approximation the transition is higher than fourth order, to the chemical potential-axis, where the transition is third order. In the second regime we use the same matrix model to obtain the deconfinement line of transitions as a function of the coupling strength and $\mu / T$ to leading order in a strong coupling expansion of lattice QCD with heavy quarks, extending previous $U(N_c)$ results to $SU(N_c)$. We show that in the case of zero chemical potential the result obtained for the Polyakov line from QCD on $S^1\times S^3$ at weak coupling reproduces the known results from the lattice strong coupling expansion, under a simple change of parameters, which is valid for sufficiently low temperatures and chemical potentials.}

\notoc
\begin{document}

\maketitle

\newpage


\section{Introduction}

Obtaining the phase diagram of QCD at non-zero chemical potential is a long-standing problem, with a picture only starting to emerge from heavy ion collisions and lattice simulations at low densities, and from models of QCD, suggesting that there is a range of phenomena at non-zero density that demands new explanations. The presence of the sign problem, however, in combination with a large coupling strength at low and moderate temperatures and chemical potentials, has the consequence that conventional techniques such as lattice simulations based on importance sampling, and infinite volume perturbation theory, are only able to access a small fraction of the overall phase diagram. To obtain a better understanding of the difficulties faced at non-zero chemical potential, and to obtain a picture of the phase diagram where conventional techniques break down, it is helpful to study limits of QCD which make calculations analytically tractable. One such limit is obtained by taking the number of colors $N_c$ large, which introduces several benefits: large $N_c$ factorization simplifies the action by removing terms that involve correlations between different points in spacetime, the infinite number of degrees of freedom provided at large $N_c$ allows for sharp phase transitions in finite volumes, and the deconfinement phase transition, which takes places at sufficiently high temperature, becomes describable in terms of the behavior of a continuous distribution of eigenvalues of the Polyakov line order parameter, which reduces the theory to an analytically solvable matrix model.

In what follows we describe in detail a calculation to obtain the Polyakov line observable and map out the confined and deconfined regions of the large $N_c$ QCD phase diagram. First, we consider QCD on $S^1 \times S^3$, where the radius, $R$, of the $S^3$ is taken sufficiently small that perturbation theory is valid, $R \ll \Lambda_{QCD}^{-1}$. In comparison with $S^1 \times {\field R}^3$, the infinite spatial volume has been sacrificed in exchange for being able to work at any temperature, $T$, where $T = \frac{1}{\beta}$ and $\beta$ is the length of the $S^1$. In
\cite{Hands:2010zp} the phase diagram for QCD with chemical potential on $S^1 \times S^3$ was calculated from various observables in the low temperature limit. In the first part of this work, we extend these results to higher temperatures to obtain the deconfinement transition line in the $\mu$-$T$ plane from the $\mu = 0$ axis to the $T = 0$ axis, where we match onto the low temperature result in \cite{Hands:2010zp}.

The differences in the weak coupling analysis in this work from \cite{Hands:2010zp} are 1) we keep the finite temperature contributions in the action, but 2) we will make the approximation that ${\boldsymbol z}_{vn} = 0$, and ${\boldsymbol z}_{fn} = 0$, for $n > 1$, where ${\boldsymbol z}_{vn}$ and ${\boldsymbol z}_{fn}$ are the single particle partition functions for vectors and fermions, for a given number of windings $n$. This approximation was also used in sections of \cite{Aharony:2003sx} to obtain concrete results at finite temperature where the corrections are exponentially suppressed at low temperatures, and power law suppressed at high temperatures. Therefore, in taking this approximation we restrict our weak-coupling results to the region of temperatures which are not too high, and we restrict the chemical potential to $\mu \lapprox \varepsilon_{f1} \big{|}_{m R \simeq 0} \simeq \frac{3}{2 R}$, where $\varepsilon_{fl} = \frac{1}{R} \sqrt{(l+\frac{1}{2})^2 + m^2 R^2}$ are the energy levels of fermions of mass $m$ formulated on $S^3$, and we consider the case where $m R \simeq 0$. This includes the region of the deconfinement line of transitions extending from the $T$ axis to the $\mu$ axis for the case of very light quarks. To obtain results for larger chemical potentials  it would be necessary to include terms with ${\boldsymbol z}_{fn} \ne 0$ for larger $n$. To consider larger quark masses it would be necessary to include terms with ${\boldsymbol z}_{vn} \ne 0$ for larger $n$. We investigate how such corrections would be included in a manner similar to that of \cite{Aharony:2003sx} and give some results for Yang-Mills theory in Section \ref{ym-corrections}.

To obtain the large $N_c$ phase diagram we follow \cite{Hands:2010zp} and generalize the technique of Gross and Witten \cite{Gross:1980he}, and Wadia \cite{Wadia:1979vk}, to include the contribution from $N_f$ quarks coupled to non-zero chemical potential, where $N_f$ is of order $N_c$. We start by reviewing Yang-Mills theory in Section \ref{ym}, and QCD with zero chemical potential in section \ref{qcd}. In each case we derive the weak coupling result on $S^1 \times S^3$ considering a more general scenario where the eigenvalues of the Polyakov line are allowed to lie off of the unit circle until the end. Then we use a simple change of parameters to show that the matrix model also reproduces the Polyakov lines for Yang-Mills theory and QCD with $\mu = 0$ from the lattice strong coupling expansion with heavy quarks in \cite{Damgaard:1986mx}. The calculations for $\mu \ne 0$ are given in Section \ref{qcd-mu}. An important feature, also imposed in \cite{Hands:2010zp}, is that the gauge fields are complexified in the case of $\mu \ne 0$. Thus we allow the Polyakov line eigenvalues to lie off of the unit circle throughout. This is crucial in obtaining the correct stationary point solution since the action of QCD with $\mu \ne 0$ is complex (This is also important in lattice simulations at finite density using stochastic quantization and complex Langevin methods. See, for example \cite{Damgaard:1987rr,Aarts:2011sf}). It is also critical to impose the $SU(N_c)$ constraint to obtain a solution which results in a $\mu$-dependent free energy and a non-zero quark number, as done in \cite{Hands:2010zp} for QCD on $S^1 \times S^3$, and in \cite{Greensite:2012xv} for effective spin models treated with mean field theory. For the $U(N_c)$ theory, at least for QCD on $S^1 \times S^3$, in effective spin models, and in the lattice strong coupling approximation \cite{Christensen:2012km}, the non-zero chemical potential results in a trivial shift of the gauge field, resulting in Polyakov lines which are given by $\rho_{\pm1} (\mu) = \rho_{\pm 1}(\mu = 0) e^{\mp \mu \beta}$.

After obtaining the Polyakov lines at weak coupling on $S^1 \times S^3$ as a function of the temperature and chemical potential and calculating the phase diagram we use the same matrix model to obtain the Polyakov lines from a lattice strong coupling expansion with heavy quarks as a function of the coupling strength and the ratio $\mu / T$, in Section \ref{lattice-results}. Our results extend the results in \cite{Damgaard:1986mx,Christensen:2012km} to $SU(N_c)$ QCD with $\mu \ne 0$ by imposing the corresponding $\sum_{i=1}^{N_c} \theta_i = 0$ constraint on the eigenvalue angles of the Polyakov line. This results in a non-zero quark number, and a $\mu$-dependent free energy. The change of parameters which makes conversion between weak and strong coupling results possible at sufficiently low temperatures and chemical potentials is derived next in Section \ref{qcd-weak-strong}. It is valid for observables such as the Polyakov lines, quark number, and free energy, which don't include correlations between different spacetime locations, in other cases the procedure would need to be generalized, if possible. It is unclear to us how it comes about that such an approximation exists which leads to the equivalence of Polyakov line matrix models connecting observables in weakly-coupled, small volume continuum QCD, to those in strongly-coupled large volume lattice QCD. It is also not clear, and would be interesting to find out, if such an approximation is also possible for other theories.

\section{1-loop QCD on $S^1 \times S^3$ vs. lattice strong coupling expansion with heavy quarks, as $N_c \rightarrow \infty$}
\label{qcd-weak-strong}

The action of large $N_c$ QCD, with large number of flavors $N_f$ and fixed $\frac{N_f}{N_c}$, to leading order in the lattice strong coupling expansion and the hopping (heavy quark) expansion is given by \cite{Damgaard:1986mx,Billo:1994ss,Christensen:2012km}
\SP{
S_{lat} - S_{Vdm} = &- J D \sum_{x} \left[ \langle W \rangle W^{\dagger}(x) + \langle W^{\dagger} \rangle W(x) - \langle W \rangle \langle W^{\dagger} \rangle \right]\\
&- h N_c \sum_{x} \left[ e^{\mu \beta} W(x) + e^{-\mu \beta} W^{\dagger}(x) \right] \, ,
\label{strong-action}
}
where $S_{Vdm}$ is the Vandermonde contribution to the action, $J \equiv 2 \left( \frac{\beta_{lat}}{2 N_c^2} \right)^{N_{\tau}}$ in terms of the inverse coupling $\beta_{lat} = \frac{2 N_c}{g^2}$, and the number of lattice sites in the temporal direction $N_{\tau}$, $h \equiv 2 \frac{N_f}{N_c} \kappa^{N_{\tau}}$ is the hopping parameter with $\kappa \equiv \frac{1}{a m+1+D}$, where $a$ is the lattice spacing, and $D$ is the number of spatial dimensions, and $W(x) = {\rm \Tr} \prod_{t=0}^{N_{\tau} - 1} U_{t,i}$ is the Polyakov line.

On $S^1 \times S^3$ the action from one loop perturbation theory takes the form \cite{Aharony:2003sx,Hands:2010zp}
\SP{
S_{S^1\times S^3} - S_{Vdm}  = &-N_c^2 \sum_{n=1}^{\infty} \frac{1}{n} {\boldsymbol z}_{vn} \rho_n \rho_{-n}\\
&+ N_f N_c \sum_{n=1}^{\infty} \frac{(- 1)^n}{n} {\boldsymbol z}_{fn} \left( e^{n \beta \mu} \rho_n + 
e^{-n \beta \mu} \rho_{-n} \right) \, ,
\label{weak-action}
}
where ${\boldsymbol z}_{vn}$ and ${\boldsymbol z}_{fn}$ are the single particle partition functions for vectors and fermions on $S^1 \times S^3$ \cite{Aharony:2003sx},
\EQ{
{\boldsymbol z}_{vn} = 2 \sum_{l=1}^{\infty} l(l+2) e^{-n \beta (l+1)/R} = \frac{2 e^{-2 n \beta / R} (3-e^{-n \beta/R})}{(1-e^{-n \beta/R})^3} \, ,
}
\EQ{
{\boldsymbol z}_{fn} = 2 \sum_{l=1}^{\infty} l (l+1) e^{-n \frac{\beta}{R} \sqrt{(l+\frac{1}{2})^2 + m^2 R^2}} \, ,
}
and the Polyakov lines are defined by $\rho_n \equiv \frac{1}{N_c} \Tr {\mathscr P} e^{n \int_{0}^{\beta} {\rm d}t A_0(x)} = \frac{1}{N_c} e^{n \beta \alpha} = \frac{1}{N_c} \sum_{i=1}^{N_c} e^{i n \theta_i}$, where ${\mathscr P}$ indicates path-ordering, and $\alpha \equiv \frac{1}{V_3} \int_{S^3} {\rm d}^3 x A_0 (x)$ with $\partial_t \alpha(t) = 0$ such that a gauge with constant, diagonal $A_0$ is chosen and massive, off-diagonal fluctuations have been integrated out. In (\ref{weak-action}) the Yang-Mills contribution has $\rho_n \rho_{-n}$ in place of $\langle W \rangle W^{\dagger}(x) + W(x) \langle W^{\dagger} \rangle - \langle W \rangle \langle W^{\dagger} \rangle$ in (\ref{strong-action}). In practice the equations of motion obtained from derivatives on $\langle W \rangle W^{\dagger}(x) + W(x) \langle W^{\dagger} \rangle - \langle W \rangle \langle W^{\dagger} \rangle$, with the expectation values fixed, are equivalent to those obtained by taking the derivatives on $\rho_n \rho_{-n}$. The actions (\ref{strong-action}) and (\ref{weak-action}) are otherwise similar with the exception of the sum over $n$ in the formulation on $S^1 \times S^3$, and that the $x$-dependence is kept in the strong coupling expansion. Due to the absence of terms with correlations between different sites in the action (\ref{strong-action}) an observable of the form $\langle F(W,W^{\dagger}) \rangle$ can be obtained as follows
\SP{
\langle F(W,W^{\dagger}) \rangle &= \frac{1}{N_x Z} \int \prod_x {\rm d}W(x) e^{-S[W(x),W^{\dagger}(x)]} \sum_{x'} F[W(x'),W^{\dagger}(x')] \, ,\\
&= \frac{\int {\rm d}W e^{-S(W,W^{\dagger})} F(W,W^{\dagger})}{\int {\rm d}W e^{-S(W,W^{\dagger})}} \, .
}
Therefore, when it is possible to approximate the sum over $n$ in (\ref{weak-action}) by the $n=1$ contribution, then it is possible to calculate observables of the form $\langle F(\rho_1,\rho_{-1}) \rangle$ in weakly-coupled QCD on $S^1 \times S^3$ and use the transformations
\SP{
\rho_1 &\leftrightarrow \frac{1}{N_c} \langle W \rangle\, ,\\
\rho_{-1} &\leftrightarrow \frac{1}{N_c} \langle W^{\dagger} \rangle \, ,\\
{\boldsymbol z}_{v1} &\leftrightarrow J D \, ,\\
{\boldsymbol z}_{f1} \frac{N_f}{N_c} &\leftrightarrow h \, ,
\label{change-vars}
}
to obtain the result for strongly-coupled lattice QCD with heavy quarks, or vice versa. The approximation of keeping only the $n=1$ contribution corresponds to taking ${\boldsymbol z}_{vn} = 0$, ${\boldsymbol z}_{fn} = 0$, for $n > 1$, which is a good approximation for $\mu < \varepsilon_{f1}$ and when the temperature is not too high (such that ${\boldsymbol z}_{v1}$, ${\boldsymbol z}_{f1} e^{\mu \beta} \gg {\boldsymbol z}_{v2}$, ${\boldsymbol z}_{f2} e^{2 \mu \beta}$) \footnote{Notice that in this limit the matrix model reduces to the one in \cite{Dumitru:2005ng}, for which the effective potential and Polyakov lines were obtained for $N_c = 3$.}. Whether the corrections, corresponding to terms including multiple windings of the Polyakov lines, can be included by considering higher order terms in the lattice strong coupling and hopping parameter expansions such that the weakly-coupled and strongly-coupled theories still share the same matrix model is left for future research.

\section{Yang-Mills theory}
\label{ym}

Taking $N_f = 0$ and evaluating the Vandermonde contribution in (\ref{weak-action}) leads to the one-loop action for Yang-Mills theory on $S^1 \times S^3$ \cite{Aharony:2003sx}
\EQ{
S_{YM} = N_c^2 \sum_{n=1}^{\infty} \frac{1}{n} \left( 1 - {\boldsymbol z}_{vn} \right) \rho_n \rho_{-n} \, .
\label{ym-action}
}
Taking the $N_c \rightarrow \infty$ limit makes it possible to define a map
\EQ{
\frac{1}{N_c} \sum_{i=1}^{N_c} \xrightarrow[N_c \rightarrow \infty]{} \int_{-\psi}^{\psi} \frac{{\rm d}s}{2 \pi} = \int_{{\cal C}} \frac{{\rm d}z}{2 \pi i} \varrho(z) \, ,
\label{sum-to-int}
}
where the $N_c$ Polyakov line eigenvalues $z_j = e^{i \theta_j}$ are distributed along the $1$-dimensional contour ${\cal C}$, which opens on the negative real axis, with endpoints at ${\tilde z} = r e^{i \psi}$, ${\tilde z}^* = r e^{-i \psi}$, where $r$ and $\psi$ are real, and the eigenvalue density is defined by
\EQ{
\varrho(z) \equiv i \frac{{\rm d}s}{{\rm d}z} \, .
}
From the map in (\ref{sum-to-int}) it is clear that the density $\varrho(z)$ is complex but evaluated along the contour ${\cal C}$ the quantity $\frac{1}{i} {\rm d}z \varrho(z)$ is real and positive and normalized as
\EQ{
\int_{{\cal C}} \frac{{\rm d} z}{2 \pi i} \varrho (z) = 1 \, .
\label{ident}
}
Note that for Yang-Mills theory the action is real and the eigenvalues $z_j$ lie on the unit circle (so $r=1$). Later, when we consider $\mu \ne 0$, the eigenvalues move off into the complex plane (the $\theta_j$ become complex), but they still lie on a $1$-dimensional arc, as we expect from the general behavior of matrix models with complex potentials, for example \cite{Dijkgraaf:2002fc}.

The map (\ref{sum-to-int}) can be used to simplify the equation of motion obtained from $\frac{\partial S}{\partial \theta_i} = 0$, which gives the stationary point solution, to the form
\EQ{
{\mathfrak P} \int_{{\cal C}} \frac{{\rm d}z'}{2 \pi i} \varrho(z') \frac{z'+z}{z'-z} = \sum_{n=1}^{\infty} {\boldsymbol z}_{vn} \left( \rho_{-n} z^n - \rho_n z^{-n} \right) \, ,
\label{eom-ym}
}
where ${\mathfrak P}$ indicates that the principal value is taken with the point $z$ left out of the range of integration, and the contour ${\cal C}$ is an arc which opens on the negative real-axis.
\subsection{Confined (ungapped) phase}

Following \cite{Gross:1980he,Wadia:1979vk} we represent the confined phase with a continuous, ungapped distribution of the Polyakov line eigenvalues, such that the contour ${\cal C}$, along which the eigenvalues are distributed, is closed and the endpoints ${\tilde z}$, ${\tilde z}^*$ coincide with $\psi = \pi$. The equation of motion (\ref{eom-ym}) can be evaluated using Cauchy's theorem. Fourier expanding the density as
\EQ{
\varrho(z) = \sum_{n=-\infty}^{\infty} \rho_n z^{-n-1} \, ,
\label{fourier-dens}
}
and collecting the residues at $z' = 0$, $z$, results in $\rho_{n} = \rho_{-n} = 0$ for $n \ne 0$. This is consistent with the result $\frac{1}{N_c} \langle W \rangle = 0$ within the confined phase from the strong coupling expansion of Yang-Mills theory. Using the identity (\ref{ident}), it is clear that $\rho_0 = 1$. From the weak coupling action (\ref{ym-action}) the confined phase persists while ${\boldsymbol z}_{v1} < 1$ \cite{Aharony:2003sx}, or, using the transformations in (\ref{change-vars}), the confined phase persists while $J D < 1$ in the case of strong coupling, as found in \cite{Damgaard:1986mx}.
\subsection{Deconfined (gapped) phase}

\begin{figure}[t]
\center
\includegraphics[width=9cm]{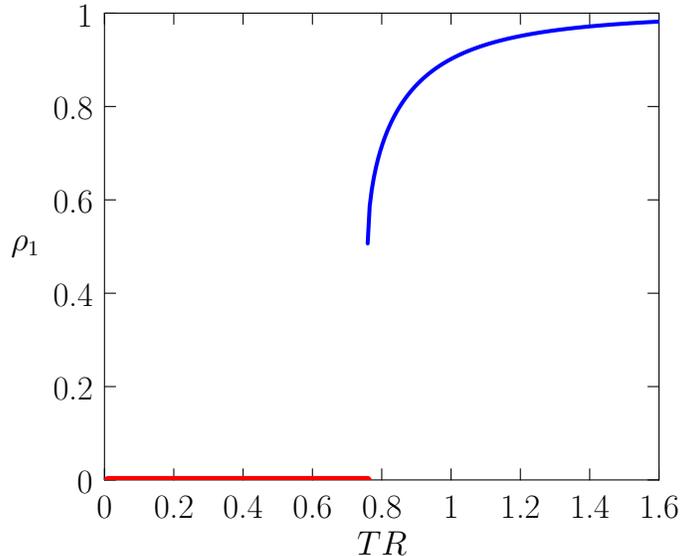}
\caption{Polyakov line $\rho_{1}$ as a function of temperature for $N_c = \infty$ Yang-Mills theory on $S^1 \times S^3$. The red line corresponds to the result from the confined phase. The blue curve corresponds to the result from the deconfined phase.}
\label{fig_dec_ym}
\end{figure}

Following \cite{Gross:1980he,Wadia:1979vk} the deconfined phase is obtained when the distribution of the Polyakov line eigenvalues develops a gap on the negative real axis, such that ${\cal C}$ is open with $\cos \psi \ne -1$. We review the result in \cite{Aharony:2003sx} for the Polyakov line from Yang-Mills theory on $S^1 \times S^3$, in the deconfined phase, leaving open the possibility that the distribution of the Polyakov line eigenvalues lies off the unit circle, and that $\rho_1 \ne \rho_{-1}$, until the end. The singular integral on the LHS of the equation of motion (\ref{eom-ym}) is solved by defining a resolvent
\EQ{
\phi (z) = \int_{{\cal C}} \frac{{\rm d} z'}{2 \pi i} \varrho (z') \frac{z'+z}{z'-z} \, .
\label{resolv}
}
Using the identity (\ref{ident}), the resolvent can be rewritten as
\EQ{
\phi (z) = \int_{{\cal C}} \frac{{\rm d}z'}{2 \pi i} \varrho(z') \frac{2 z'}{z'-z} - 1 \, .
}
Therefore $\phi(z)$ satisfies the Plemelj formulae
\EQ{
\phi^+(z) - \phi^-(z) = 2 z \varrho(z) \, ,
\label{ple1}
}
\EQ{
\phi^+(z) + \phi^-(z) = 2 \sum_{n=1}^{\infty} {\boldsymbol z}_{vn} \left( \rho_{-n} z^n - \rho_n z^{-n} \right) \, ,
\label{ple2}
}
by taking the contour ${\cal C}$ to lie on a square root branch cut. Following \cite{Wadia:1979vk} the Plemelj formulae are solved writing
\EQ{
\phi(z) = h(z) H(z) \, ,
\label{decomp}
}
with
\EQ{
h(z) = \sqrt{(z - {\tilde z}) (z - {\tilde z}^*)} \, .
}
From (\ref{ple2})
\EQ{
H(z) = \frac{1}{2 \pi i} \oint_{\Gamma} \frac{{\rm d}z'}{z'-z} \frac{\sum_{n=1}^{\infty} {\boldsymbol z}_{vn} (\rho_{-n} z'^n - \rho_n z'^{-n})}{\sqrt{(z'-{\tilde z})(z'-{\tilde z}^*)}} \, ,
}
where $\Gamma$ is a contour around ${\cal C}$ which can be peeled off to evaluate the integral by collecting the residues at $0$, $\infty$, and $z$ to obtain
\SP{
H(z) = &\left[ (z - {\tilde z}) (z - {\tilde z}^*) \right]^{-1/2} \sum_{n=1}^{\infty} {\boldsymbol z}_{vn} \left( \rho_{-n} z^n - \rho_n z^{-n} \right)\\
&+ \sum_{l=1}^{\infty} \sum_{k=0}^{\infty} P_k (\cos \psi) {\boldsymbol z}_{v(l+k)} \left( \rho_{l+k} r^{-k-1} z^{-l} + \rho_{-l-k} r^k z^{l-1} \right) \, .
}
From (\ref{decomp}) this results in
\SP{
\phi (z) = &\sum_{n=1}^{\infty} {\boldsymbol z}_{vn} \left( \rho_{-n} z^n - \rho_n z^{-n} \right) + \sqrt{(z - {\tilde z}) (z - {\tilde z}^*)}\\
&\times \sum_{l=1}^{\infty} \sum_{k=0}^{\infty} P_k (\cos \psi) {\boldsymbol z}_{v(l+k)} \left( \rho_{l+k} r^{-k-1} z^{-l} + \rho_{-l-k} r^k z^{l-1} \right) \, ,
\label{res-ym}
}
which agrees with \cite{Jurkiewicz:1982iz}, where $P_k (x)$ are the Legendre Polynomials. The density is obtained from (\ref{ple1}) as
\SP{
\varrho(z) = &\sqrt{(z - {\tilde z}) (z - {\tilde z}^*)}\\
&\times \sum_{l=1}^{\infty} \sum_{k=0}^{\infty} P_k (\cos \psi) {\boldsymbol z}_{v(l+k)} \left( \rho_{l+k} r^{-k-1} z^{-l-1} + \rho_{-l-k} r^k z^{l-2} \right) \, .
}
The Polyakov lines can be calculated from
\EQ{
\rho_n = \int_{\cal C} \frac{{\rm d}z}{2 \pi i} \varrho(z) z^n \, ,
\label{polys}
}
where again we write the integration along ${\cal C}$ as the contour $\Gamma$, then peel $\Gamma$ off and evaluate the integral collecting the residues outside. Integrating both sides of (\ref{ple1}) one obtains the more general transformation
\EQ{
\int_{{\cal C}} \frac{{\rm d}z}{2 \pi i} \varrho(z) F(z) = \oint_{\Gamma} \frac{{\rm d}z}{4 \pi i z} \phi(z) F(z) \, ,
}
such that the Polyakov lines can be obtained from
\EQ{
\rho_n = \oint_{\Gamma} \frac{{\rm d}z}{4 \pi i z} \phi(z) z^{n} \, ,
\label{rho-ym-dec}
}
where $\rho_0$ is the identity (\ref{ident}). If we make the approximation that ${\boldsymbol z}_{vn} = 0$ for $n > 1$, then
\EQ{
\rho_0 = 1 = \frac{1}{2} {\boldsymbol z}_{v1} \left( \rho_1 r^{-1} + \rho_{-1} r \right) \left( 1 - \cos \psi \right) \, ,
}
gives the constraint
\EQ{
\rho_1 r^{-1} + \rho_{-1} r = \frac{2}{{\boldsymbol z}_{v1}(1-\cos \psi)} \, .
\label{ym_constr1}
}
Similarly one can solve (\ref{rho-ym-dec}) for $n = \pm 1$ to obtain
\EQ{
\rho_1 = \frac{\frac{1}{2}r^2 {\boldsymbol z}_{v1}\rho_{-1}\sin^{2}\psi}{2-{\boldsymbol z}_{v1}(1-\cos\psi)} \, ,
}
\EQ{
\rho_{-1} = \frac{\frac{1}{2}r^2 {\boldsymbol z}_{v1}\rho_{1}\sin^{2}\psi}{2-{\boldsymbol z}_{v1}(1-\cos\psi)} \, ,
}
which imply the constraint
\EQ{
2 - {\boldsymbol z}_{v1}(1-\cos \psi) = \frac{1}{2} {\boldsymbol z}_{v1} \sin^2 \psi \, .
}
The solution from this constraint is
\EQ{
\cos \psi = -1 + \frac{2}{{\boldsymbol z}_{v1}} \sqrt{{\boldsymbol z}_{v1}^2-{\boldsymbol z}_{v1}} \, ,
}
such that (\ref{ym_constr1}) gives an expression for the Polyakov lines as a function of the temperature
\EQ{
\rho_1 r^{-1} + \rho_{-1} r = 1 + \frac{1}{{\boldsymbol z}_{v1}} \sqrt{{\boldsymbol z}_{v1}^2-{\boldsymbol z}_{v1}} \, .
}
For Yang-Mills theory the action is real so we take $r = 1$ and $\rho_{-1} = \rho_1$ such that the Polyakov lines simplify to
\EQ{
\rho_1 = \rho_{-1} = \frac{1}{2} \left[ 1 + \frac{1}{{\boldsymbol z}_{v1}} \sqrt{{\boldsymbol z}_{v1}^2-{\boldsymbol z}_{v1}} \right] \, ,
\label{rho-dec-ym}
}
which is the result in \cite{Aharony:2003sx} and we plot it in Figure \ref{fig_dec_ym} along with the $\rho_{\pm 1} = 0$ result for the confined phase from the previous subsection to illustrate the discontinuity at the transition temperature. The transition occurs when ${\boldsymbol z}_{v1}=1$ which corresponds to $T R \approx 0.759$ where $\rho_{\pm 1}$ jumps from $0$ to $\frac{1}{2}$. As $ T R \rightarrow \infty$, ${\boldsymbol z}_{v1} \rightarrow \infty$ and $\rho_{\pm 1} \rightarrow 1$. It is interesting to note that there are lattice simulations of Yang-Mills theory with $N_c = 4$ and $5$ which also show a jump in the (renormalized) Polyakov line from $0$ to a value close to $\frac{1}{2}$ \cite{Mykkanen:2012ri} \footnote{The value to which the renormalized Polyakov line jumps in lattice simulations is generally a scheme-dependent quantity; the scheme chosen in \cite{Mykkanen:2012ri} is based on the assumption that the renormalized zero-temperature quark-antiquark potential does not include a term independent of the distance. We thank Marco Panero for bringing this research to our attention.}.

Notice that under the change of parameters (\ref{change-vars}), (\ref{rho-dec-ym}) becomes the known result for Yang-Mills theory from the lattice strong coupling expansion \cite{Damgaard:1986mx}
\EQ{
\frac{1}{N_c} \langle W \rangle = \frac{1}{2} \left[ 1 + \sqrt{1 - \frac{1}{J D}} \right] \, ,
}
where the transition point ${\boldsymbol z}_{v1} = 1$ is converted to $J D = 1$, as in \cite{Damgaard:1986mx}.
\subsubsection{Corrections}
\label{ym-corrections}

\begin{figure}[t]
\center
\includegraphics[width=9cm]{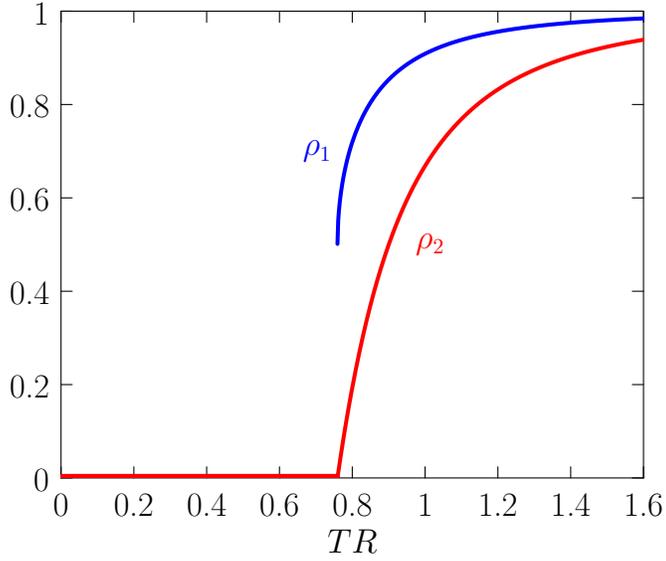}
\caption{Polyakov lines $\rho_{1}$, $\rho_2$ as a function of temperature for $N_c = \infty$ Yang-Mills theory (including contributions resulting from ${\boldsymbol z}_{vn} \ne 0$ for $n = 1$, $2$.). The blue curve corresponds to $\rho_1$ (note that there is also a disconnected section of the blue curve along the $T R$ axis until ${\boldsymbol z}_{v1} = 1$). The red curve corresponds to $\rho_2$.}
\label{fig-trp12}
\end{figure}

It is straightforward to consider the ${\boldsymbol z}_{v2} \ne 0$ corrections to the Polyakov lines for Yang-Mills theory in the deconfined phase. In this section we simplify our notation by taking $x \equiv \cos \psi$. The identity constraint given by (\ref{rho-ym-dec}) for $n = 0$ becomes
\EQ{
1 = \frac{1}{4} (1-x) \left[ 2 {\boldsymbol z}_{v1} (\rho_1 r^{-1} + \rho_{-1} r) + {\boldsymbol z}_{v2} (1 + 3 x) (\rho_2 r^{-2} + \rho_{-2} r^2) \right] \, .
}
Because the action is real we take $r=1$ which leads to $\rho_n = \rho_{-n}$. (\ref{rho-ym-dec}) for $n = 1$, $2$, gives
\EQ{
\rho_1 = \frac{4(1+x)^2}{4(1+3 x)+{\boldsymbol z}_{v1} (1-x)^3} \, ,
}
\EQ{
\rho_2 = \frac{8 - 2 {\boldsymbol z}_{v1} (1-x)(3+x)}{(1-x) {\boldsymbol z}_{v2} \left[ 4 (1 + 3 x) + {\boldsymbol z}_{v1} (1-x)^3 \right]} \, ,
}
where $x \equiv \cos \psi$ is the solution of
\EQ{
1 = \frac{8(1-x)^2 (1+x)^4 {\boldsymbol z}_{v1} {\boldsymbol z}_{v2}}{\left[ 4 - {\boldsymbol z}_{v1} (1-x) (3+x) \right] \left[ 16 + {\boldsymbol z}_{v2} (1-x) \left[ 7 + x \left( 3 + x (13 + 9 x) \right) \right] \right] }
}
which is real and has $\left| x \right| \le 1$. The correct root depends on $\frac{\beta}{R}$: it changes when ${\boldsymbol z}_{v2} = 1$, corresponding to $T R \approx 1.519$. However, the Polyakov lines $\rho_1$, $\rho_2$ remain continuous over $T R \approx 1.519$, as shown in Figure \ref{fig-trp12}. Notice that $\rho_1$ in Figure \ref{fig-trp12}, which contains the ${\boldsymbol z}_{v2} \ne 0$ corrections, has only barely changed from the result with ${\boldsymbol z}_{v2} = 0$ in Figure \ref{fig_dec_ym}.

\section{QCD with $\mu = 0$}
\label{qcd}

The one-loop action for QCD with $\mu = 0$ is \cite{Aharony:2003sx}
\EQ{
S_{QCD} = N_c^2 \sum_{n=1}^{\infty} \frac{1}{n} \left( 1 - {\boldsymbol z}_{vn} \right) \rho_n \rho_{-n} + N_f N_c \sum_{n=1}^{\infty} \frac{(-1)^n}{n} {\boldsymbol z}_{fn} \left( \rho_n + \rho_{-n} \right) \, ,
\label{Sqcdmu0}
}
such that the equation of motion obtained from $\frac{\partial S}{\partial \theta_i} = 0$ becomes
\SP{
{\mathfrak P} \int_{{\cal C}} \frac{{\rm d}z'}{2 \pi i} \varrho(z') \frac{z'+z}{z'-z} = \sum_{n=1}^{\infty} \bigg[ &\left( {\boldsymbol z}_{vn} \rho_{-n} - \frac{N_f}{N_c} (-1)^n {\boldsymbol z}_{fn} \right) z^n\\
&- \left( {\boldsymbol z}_{vn} \rho_n - \frac{N_f}{N_c} (-1)^n {\boldsymbol z}_{fn} \right) z^{-n} \bigg] \, ,
\label{eom-qcd-mu0}
}
which corresponds to the Yang-Mills equation of motion (\ref{eom-ym}) under the shift ${\boldsymbol z}_{vn} \rho_{\pm n} \rightarrow {\boldsymbol z}_{vn} \rho_{\pm n} - \frac{N_f}{N_c} (-1)^n {\boldsymbol z}_{fn}$.
\subsection{Confined phase}

As in Yang-Mills theory the contour ${\cal C}$ is closed in the confined phase and the equation of motion (\ref{eom-qcd-mu0}) is evaluated by Fourier expanding the density as in (\ref{fourier-dens}) and collecting the residues at $0$, $z$, resulting in
\EQ{
\rho_{\pm n} = \frac{N_f}{N_c} \frac{(-1)^{n+1}{\boldsymbol z}_{fn}}{1-{\boldsymbol z}_{vn}} \, ,
\label{rho-n-con-qcd0}
}
for $n \ne 0$, in agreement with \cite{Schnitzer:2004qt}. For $n = 0$, $\rho_0 = 1$ as in Yang-Mills theory. For $n=1$ the lattice strong coupling result can be obtained by using the transformations in (\ref{change-vars}) to give
\EQ{
\frac{1}{N_c} \langle W \rangle = \frac{h}{1 - J D} \, ,
}
in agreement with \cite{Damgaard:1986mx}.

There is an upper limit on the Polyakov line that can be obtained while the distribution remains ungapped. A consequence of working in the approximation where ${\boldsymbol z}_{vn} = {\boldsymbol z}_{fn} = 0$ for $n > 1$, is that the Polyakov lines, $\rho_{n} = 0$ for $n > 1$ in the confined phase. The eigenvalue density (\ref{fourier-dens}) obtains a zero,
\EQ{
\varrho(z) = \rho_{-1} + z^{-1} + \rho_1 z^{-2} = 0 \, ,
}
when
\EQ{
z = \frac{-1 \pm \sqrt{1 - 4 \rho_1 \rho_{-1}}}{2 \rho_{-1}}
}
lies on the Polyakov line eigenvalue distribution, on ${\cal C}$. Since the distribution lies on the unit circle when $\mu = 0$, with $\rho_{-n} = \rho_{n}$, the gap forms at $z = -1$, which corresponds to $\rho_1 = \rho_{-1} = \frac{1}{2}$. This gives the highest temperature at which the ungapped phase can end. Plugging $\rho_{\pm 1} = \frac{1}{2}$ into (\ref{rho-n-con-qcd0}) gives the allowed region of the ungapped phase as
\EQ{
1 - {\boldsymbol z}_{v1} - 2 {\boldsymbol z}_{f1} \frac{N_f}{N_c} \ge 0 \, .
\label{qcd0-Tc-con}
}
\subsection{Deconfined phase}

After a gap forms in the Polyakov line eigenvalue distribution the theory enters the deconfined phase, where the contour ${\cal C}$ lies on an arc which opens up on the negative real-axis. Comparison of the equation of motion (\ref{eom-qcd-mu0}) with the equation of motion for Yang-Mills theory (\ref{eom-ym}) indicates that there are no new poles in $z$ so the results obtained in the deconfined phase for Yang-Mills theory should carry over to QCD with $\mu = 0$ by shifting ${\boldsymbol z}_{vn} \rho_{\pm n} \rightarrow {\boldsymbol z}_{vn} \rho_{\pm n} - \frac{N_f}{N_c} (-1)^n {\boldsymbol z}_{fn}$.

Taking ${\boldsymbol z}_{vn} = {\boldsymbol z}_{fn} = 0$ for $n > 1$, and defining $x \equiv \cos \psi$, the constraint from the identity, (\ref{rho-ym-dec}) with $n = 0$, becomes
\SP{
\rho_0 = 1 = ~&\frac{1}{2} \left( 1 - x \right) \left[ \left( {\boldsymbol z}_{v1} \rho_1 +\frac{N_f}{N_c} {\boldsymbol z}_{f1} \right) r^{-1} + \left( {\boldsymbol z}_{v1} \rho_{-1} +\frac{N_f}{N_c} {\boldsymbol z}_{f1} \right) r \right] \, .
\label{constr1-qcd}
}
For $n = \pm 1$ (\ref{rho-ym-dec}) gives the Polyakov lines
\EQ{
\rho_{\pm 1} = \frac{{\boldsymbol z}_{f1} \frac{N_f}{N_c} (1-x) \left[ 8+4 r^{\pm 2} (1+x)- {\boldsymbol z}_{v1} (1-x)^2 (3+x)\right]}{16 - {\boldsymbol z}_{v1} (1-x) \left[16 - {\boldsymbol z}_{v1} (1-x)^2 (3+x) \right]} \, .
}
Since $\mu = 0$ the action is real and the eigenvalues of the Polyakov lines lie on the unit circle with $r=1$, resulting in
\EQ{
\rho_{1} = \rho_{-1} = \frac{(1-x)(3+x){\boldsymbol z}_{f1} \frac{N_f}{N_c}}{4-(1-x)(3+x){\boldsymbol z}_{v1}} \, .
\label{rho-qcd-r1}
}
Solving the constraint (\ref{constr1-qcd}) for $x$ gives
\EQ{
x = \frac{-{\boldsymbol z}_{v1} - 2 {\boldsymbol z}_{f1} \frac{N_f}{N_c} + 2 \sqrt{{\boldsymbol z}_{v1}^2 - {\boldsymbol z}_{v1} + 2 {\boldsymbol z}_{v1} {\boldsymbol z}_{f1} \frac{N_f}{N_c} + {\boldsymbol z}_{f1}^2 \frac{N_f^2}{N_c^2}}}{{\boldsymbol z}_{v1}} \, ,
\label{xeq-mu0}
}
where we have chosen the root that gives $\left| x \right| \le 1$ as a function of $T$. Plugging this into (\ref{rho-qcd-r1}), allows us to solve for $\rho_{\pm 1}$ as a function of $T$,
\EQ{
\rho_{\pm 1} = \frac{{\boldsymbol z}_{v1} - {\boldsymbol z}_{f1} \frac{N_f}{N_c} + \sqrt{{\boldsymbol z}_{v1}^2 - {\boldsymbol z}_{v1} + 2 {\boldsymbol z}_{v1} {\boldsymbol z}_{f1} \frac{N_f}{N_c} + {\boldsymbol z}_{f1}^2 \frac{N_f^2}{N_c^2}}}{2 {\boldsymbol z}_{v1}} \, ,
\label{rho-dec-qcd0}
}
which is the result in \cite{Schnitzer:2004qt}. This reduces to the Yang-Mills result (\ref{rho-dec-ym}) when $\frac{N_f}{N_c} \rightarrow 0$. It also matches onto the result in \cite{Damgaard:1986mx} for $\mu = 0$ under the transformations (\ref{change-vars}), that is
\EQ{
\frac{1}{N_c} \langle W \rangle = \frac{1}{2} \left( 1 - \frac{h}{J D} \right) + \frac{1}{2} \sqrt{\left( 1 - \frac{h}{J D} \right)^2 - \frac{1}{J D} \left( 1 - 4 h \right)} \, .
}
The equation for $x(T)$ in (\ref{xeq-mu0}) only has a solution with $\left| x \right| \le 1$ for
\EQ{
1 - {\boldsymbol z}_{v1} - 2 {\boldsymbol z}_{f1} \frac{N_f}{N_c} \le 0 \, .
}
The minimum temperature occurs when the LHS=0. From (\ref{qcd0-Tc-con}) this is precisely the temperature at which the confined phase must end and plugging it into (\ref{rho-dec-qcd0}) results in $\rho_{\pm 1} = \frac{1}{2}$. Therefore the transition is smooth and occurs at precisely
\EQ{
1 - {\boldsymbol z}_{v1} - 2 {\boldsymbol z}_{f1} \frac{N_f}{N_c} = 0 \, ,
\label{qcd0-Tc}
}
as in \cite{Schnitzer:2004qt,Basu:2008uc}, or in the case of the strong coupling expansion, using the transformations in (\ref{change-vars}), the transition occurs when
\EQ{
1 - J D - 2 h = 0 \, ,
}
in agreement with \cite{Damgaard:1986mx}.

\begin{figure}[t]
\center
\includegraphics[width=9cm]{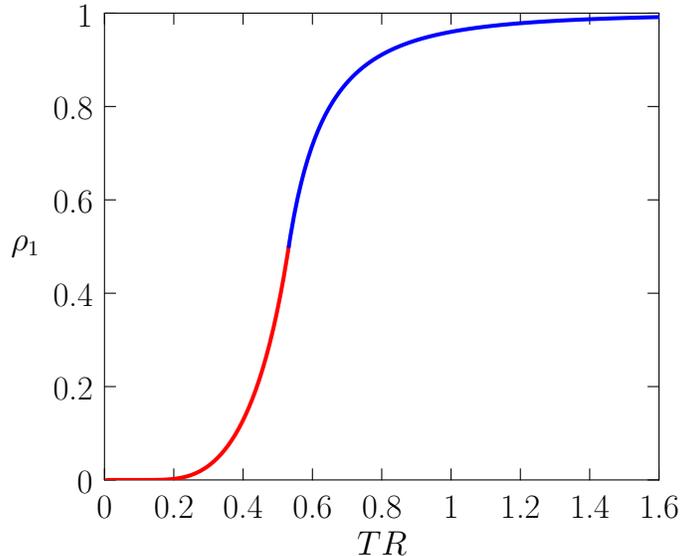}
\caption{Polyakov line $\rho_{1}$ as a function of temperature for QCD with $N_c$, $N_f = \infty$, $\frac{N_f}{N_c} = 1$, and $m R = 0$. The red curve corresponds to the ungapped phase. The blue curve corresponds to the gapped phase.}
\label{fig_dec_qcd0}
\end{figure}

The weak-coupling result for the Polyakov line (\ref{rho-dec-qcd0}) is plotted in Figure \ref{fig_dec_qcd0} for $m R = 0$, along with the result in confined phase (\ref{rho-n-con-qcd0}) from the previous subsection, where the transition point occurs at the temperature given by (\ref{qcd0-Tc}), corresponding to $\rho_{1} = \frac{1}{2}$. This transition is at least fifth order, and could potentially be infinite order, resulting in a crossover, since a pattern appears to emerge in the derivatives of the partition function with respect to the temperature, evaluated at the transition point \footnote{We would like to thank Andr{\'a}s Patk{\'o}s for bringing this possibility to our attention.}. We have found that
\EQ{
\frac{1}{N_c^2} \frac{\partial^k (\log Z)}{\partial T^k} = \frac{1}{2} \frac{N_f}{N_c} {\boldsymbol z}_{f1}^{(k)} \, ,
\label{logZderivs}
}
where $k$ refers to the number of derivatives with respect to $T$, and we have checked that this is true up to $k = 4$. Here, $\log Z$ is obtained from (\ref{Sqcdmu0}) in the limit ${\boldsymbol z}_{vn} = {\boldsymbol z}_{fn} = 0$ for $n > 1$, and using $S = -\log Z$, valid at the saddle point. (\ref{logZderivs}) holds in both the confined and deconfined phases, where $\log Z$ is evaluated with $\rho_1$ obtained from (\ref{rho-n-con-qcd0}) and (\ref{rho-dec-qcd0}), respectively. The resulting $\frac{\partial^k (\log Z)}{\partial T^k}$ in each phase is evaluated transition point (\ref{qcd0-Tc}), satisfying the derivatives of (\ref{qcd0-Tc}) as well. For the lattice theory (\ref{logZderivs}) becomes
\EQ{
\frac{1}{N_c^2} \frac{\partial^k (\log Z)}{\partial T^k} = \frac{1}{2} h^{(k)} \, .
}
It is unclear how the order of the transition would be affected when corrections are taken into account from ${\boldsymbol z}_{vn}$ and ${\boldsymbol z}_{fn}$ for $n > 1$ for the theory on $S^1 \times S^3$, or additional terms in the strong coupling and hopping parameter expansions for the lattice theory. This is a possible direction for future research.

It would be interesting to see what happens to the order of the transition as the quark mass is varied. As $m \rightarrow \infty$ the quark contribution becomes less significant and the theory approaches pure Yang-Mills. To accurately determine the order of the transition for the theory on $S^1 \times S^3$ for larger quark masses, whether the transition remains continuous or not to high order, it would be necessary to include corrections from ${\boldsymbol z}_{vn} \ne 0$ for $n > 1$, which, one by one, become of the same order of $z_{f1}$ when the mass is increased. In the strongly coupled lattice theory this corresponds to including corrections to the leading order in the hopping expansion.

We know from the low temperature results for $\mu \gapprox \varepsilon_{f1}$ in \cite{Hands:2010zp} that the transition becomes third order in the limit $\mu \rightarrow \varepsilon_{f1}$. In what follows we expect a sharpening of the transition for the theory on $S^1 \times S^3$ as a function of $T R$ for increasing $\mu R$, or in the lattice theory as a function of $J$ for increasing $\mu/T$, but it is difficult to obtain the order of the transition between the $\mu = 0$, and $T = 0$ (or $J = 0$) endpoints, because the derivatives of $\log Z$ become complicated and need to be solved numerically. We are only be able to determine that the order of the line of transitions in the region $0 < \mu < \varepsilon_{f1}$ is at least second order from the continuity of the quark number across the transition.

\section{QCD with $\mu < \varepsilon_{f1}$}
\label{qcd-mu}

The one-loop action for QCD with $\mu \ne 0$ is \cite{Hands:2010zp}
\SP{
S_{QCD} = &N_c^2 \sum_{n=1}^{\infty} \frac{1}{n} \left( 1 - {\boldsymbol z}_{vn} \right) \rho_n \rho_{-n}\\
&+ N_f N_c \sum_{n=1}^{\infty} \frac{(-1)^n}{n} {\boldsymbol z}_{fn} \left( \rho_n e^{n \mu \beta} + \rho_{-n} e^{-n \mu \beta} \right) + i {\cal N} N_c \sum_{i=1}^{N_c} \theta_i \, ,
}
where ${\cal N}$ is a Lagrange multiplier which imposes the $SU(N_c)$ constraint $\sum_{i=1}^{N_c} \theta_i = 0$. The equation of motion from $\frac{\partial S}{\partial \theta_i} = 0$ becomes
\EQ{
{\mathfrak P} \int_{{\cal C}} \frac{{\rm d}z'}{2 \pi i} \varrho(z') \frac{z'+z}{z'-z} = \sum_{n=1}^{\infty} \left( \alpha_{-n} z^n - \alpha_n z^{-n} \right) - {\cal N} \, ,
\label{eom-mu}
}
where we have defined $\alpha_{\pm n} \equiv {\boldsymbol z}_{vn} \rho_{\pm n} - \frac{N_f}{N_c} (-1)^n {\boldsymbol z}_{fn} e^{\mp n \mu \beta}$. This reduces to (\ref{eom-qcd-mu0}) when $\mu = 0$, and (\ref{eom-ym}) when $N_f = 0$, where we will find that ${\cal N} = 0$ in the confined phase, and when $r = 1$ in the deconfined phase.
\subsection{Confined phase}

In the region of the confined phase with $\mu < \varepsilon_{f1}$ solving the equation of motion (\ref{eom-mu}) with the density (\ref{fourier-dens}) gives
\EQ{
\rho_{\pm n} = \frac{N_f}{N_c} \frac{(-1)^{n+1}{\boldsymbol z}_{fn} e^{\mp n \mu \beta}}{1-{\boldsymbol z}_{vn}} \, .
\label{rho-n-con-qcd1}
}
and ${\cal N} = 0$ in agreement with \cite{Hands:2010zp}. Considering $n = 1$ and making the replacements in (\ref{change-vars}) gives the results for the lattice strong coupling theory
\EQ{
\frac{1}{N_c} \langle W \rangle = \frac{h e^{-\mu \beta}}{1 - J D} \, ,
}
\EQ{
\frac{1}{N_c} \langle W^{\dagger} \rangle = \frac{h e^{\mu \beta}}{1 - J D} \, ,
}
in agreement with \cite{Christensen:2012km}.
\subsection{Deconfined phase}

The lack of new poles in the equation of motion (\ref{eom-mu}) resulting from the contribution of quarks with $\mu \ne 0$ means that the resolvent takes the form in (\ref{res-ym}) with the shift ${\boldsymbol z}_{vn} \rho_{\pm n} \rightarrow \alpha_{\pm n}$, such that it becomes
\SP{
\phi (z) = &- {\cal N} + \sum_{n=1}^{\infty} \left( \alpha_{-n} z^n - \alpha_n z^{-n} \right) + \sqrt{(z - {\tilde z}) (z - {\tilde z}^*)}\\
&\times \sum_{l=1}^{\infty} \sum_{k=0}^{\infty} P_k (\cos \psi) \left( \alpha_{l+k} r^{-k-1} z^{-l} + \alpha_{-l-k} r^k z^{l-1} \right) \, .
\label{res-mu1}
}
Using (\ref{rho-ym-dec}) with $n = 0$ and working in the limit ${\boldsymbol z}_{vn} = {\boldsymbol z}_{fn} = 0$ for $n > 1$ we obtain
\EQ{
\rho_0 = 1 = \frac{1}{2} \left( 1 - x \right) \left( \alpha_1 r^{-1} + \alpha_{-1} r \right) \, .
\label{ident1-qcd1}
}
The Polyakov lines obtained from (\ref{rho-ym-dec}) with $n = \pm 1$ are
\EQ{
\rho_{\pm 1} = \frac{{\boldsymbol z}_{f1} \frac{N_f}{N_c} (1-x) e^{\mp \mu \beta} \left[ 8+4 e^{\pm 2 \mu \beta} r^{\pm 2} (1+x)- {\boldsymbol z}_{v1} (1-x)^2 (3+x)\right]}{16 - {\boldsymbol z}_{v1} (1-x) \left[16 - {\boldsymbol z}_{v1} (1-x)^2 (3+x) \right]} \, .
\label{polys-qcd1-1}
}
Plugging (\ref{polys-qcd1-1}) into (\ref{ident1-qcd1}) results in a relationship between $r$ and $x$ given by
\EQ{
1 = \frac{2 {\boldsymbol z}_{f1} \frac{N_f}{N_c}(1-x)(1+r^2 e^{2 \mu \beta})}{r e^{\mu \beta}\left[4 - {\boldsymbol z}_{v1} (1-x)(3+x) \right]} \, .
\label{xr-eq1}
}
The solution for $r$ which allows for matching onto the $\mu = 0$ result is given by the positive root,
\EQ{
r = \frac{4 - {\boldsymbol z}_{v1} (1-x)(3+x) + \sqrt{\left[ 4 - {\boldsymbol z}_{v1} (1-x)(3+x) \right]^2 - 16 (1-x)^2 {\boldsymbol z}_{f1}^2 \frac{N_f^2}{N_c^2}}}{4 (1-x) e^{\mu \beta} {\boldsymbol z}_{f1} \frac{N_f}{N_c}} \, .
\label{r-eq-mu}
}
Another two constraints are necessary to obtain $x$ or $r$ as a function of the temperature (two because there is also a dependence of $x$ and $r$ on ${\cal N}$). One constraint is that the solution must satisfy \cite{Muskhelishvili:2008si,Jurkiewicz:1982iz}
\EQ{
0 = \int_{{\cal C}} \frac{{\rm d}z}{2 \pi i} \frac{\sum_{n=1}^{\infty} \left( \alpha_{-n} z^n - \alpha_n z^{-n} \right) - {\cal N} + 1}{\sqrt{(z-{\tilde z})(z-{\tilde z}^*)}} \, ,
\label{muskhelish}
}
which can be evaluated to obtain
\EQ{
\sum_{k=0}^{\infty} \alpha_{k+1} P_{k}(x) r^{-(k+1)} = 1 - {\cal N} + \sum_{k=1}^{\infty} \alpha_{-k} P_k (x) r^k \, .
\label{ident2}
}
This agrees with the form in \cite{Jurkiewicz:1982iz,Aharony:2003sx} for ${\cal N} = 0$, $N_f = 0$ and $r=1$. At least for the concerned case with ${\boldsymbol z}_{vn} = {\boldsymbol z}_{fn} = 0$ for $n > 1$ this constraint is equivalent to that which results from expanding the resolvent in the $z \rightarrow \infty$ limit. The $z \rightarrow 0$ and $z \rightarrow \infty$ expansions can be performed by using (\ref{resolv}), (\ref{ident}), and (\ref{polys}), and these take the form
\EQ{
\lim_{z \rightarrow 0} \phi(z) = 1 + 2 \sum_{n=1}^{\infty} z^n \rho_{-n} \, ,
\label{res-0}
}
\EQ{
\lim_{z \rightarrow \infty} \phi(z) = -1 -2 \sum_{n=1}^{\infty} \frac{1}{z^n} \rho_n \, .
\label{res-inf}
}
The $z^0$ terms in (\ref{res-0}), (\ref{res-inf}) can be matched against the corresponding terms of the $z \rightarrow 0,\infty$ expansions of (\ref{res-mu1}) to obtain the constraints
\EQ{
1 = -{\cal N} - x \alpha_1 r^{-1} + \alpha_{-1} r \, ,
\label{con-res-0}
}
\EQ{
-1 = -{\cal N} - \alpha_1 r^{-1} + x \alpha_{-1} r \, .
\label{con-res-inf}
}
Notice that (\ref{con-res-inf}) gives the same constraint as (\ref{ident2}). Also the $z^{\pm 1}$ constraints from (\ref{res-0}), (\ref{res-inf}) reproduce the Polyakov lines in (\ref{polys-qcd1-1}). The constraint (\ref{con-res-0}) could also be obtained using the identity constraint (\ref{ident1-qcd1}) with (\ref{con-res-inf}). Using (\ref{polys-qcd1-1}) in (\ref{con-res-0}), (\ref{con-res-inf}) and solving for ${\cal N}$ gives
\EQ{
{\cal N} = (1+x) \left[ -\frac{1}{1-x} + \frac{4 {\boldsymbol z}_{f1} \frac{N_f}{N_c} r^{-1} e^{-\mu \beta} \left[ {\boldsymbol z}_{v1} - {\boldsymbol z}_{v1} x^2 +2 r^2 e^{2 \mu \beta} (2 - {\boldsymbol z}_{v1} (1-x)) \right]}{16 - {\boldsymbol z}_{v1} (1-x) \left[ 16 - {\boldsymbol z}_{v1} (1-x)^2 (3+x) \right]} \right] \, ,
\label{n-quark}
}
where the equation for $x(T R)$ obtained by plugging $r$ from (\ref{r-eq-mu}) into (\ref{n-quark}) reduces to that of QCD with $\mu = 0$, if ${\cal N} = 0$, as expected. For $\mu \ne 0$ it becomes possible to have ${\cal N} \ne 0$ because it corresponds to the normalized quark number ${\cal N} = \frac{1}{N_c^2} N_q$ as in \cite{Hands:2010zp}. This can be seen by summing (\ref{eom-mu}) over all the eigenvalues, $z_j = e^{i \theta_j}$, of the Polyakov line, and solving for the Lagrange multiplier to obtain
\SP{
{\cal N} &= \frac{1}{N_c} \sum_{i=1}^{N_c} \sum_{n=1}^{\infty} \left( \alpha_{-n} z_i^n - \alpha_{n} z_i^{-n} \right)\\
&= \frac{N_f}{N_c^2} \sum_{i=1}^{N_c} \sum_{n=1}^{\infty} (-1)^n {\boldsymbol z}_{fn} \left( e^{n \mu \beta} z_i^n - e^{-n \mu \beta} z_i^{-n} \right)\\
&= \frac{N_f}{N_c^2} \sum_{i=1}^{N_c} \sum_{l=1}^{\infty} 2 l(l+1) \left[ \frac{1}{1 + e^{\beta \varepsilon_{fl} - i \theta_i - \mu \beta}} - \frac{1}{1 + e^{\beta \varepsilon_{fl} + i \theta_i + \mu \beta}} \right]\\
&\xrightarrow[N_c \rightarrow \infty]{} \frac{1}{N_c^2} \frac{1}{\beta} \frac{\partial \log Z}{\partial \mu} \, .
}
Since (\ref{muskhelish}) - (\ref{con-res-inf}) have only introduced one new constraint it is necessary to impose an additional constraint to solve for ${\cal N}$, the $SU(N_c)$ constraint. In the large $N_c$ limit $\sum_{i=1}^{N_c} \theta_i = 0$ takes the form
\EQ{
\int_{{\cal C}} \frac{{\rm d}z}{2 \pi i} \varrho(z) \log(z) = 0 \, .
}
Defining the contour $\Gamma$ around ${\cal C}$ and peeling it off to enclose the poles at $0$ and $\infty$, and the branch cut of $\log z$ on the negative real axis, gives
\SP{
\oint_{\Gamma} \frac{{\rm d}z}{2 \pi i z} \phi(z) \log(z) &= \lim_{\epsilon \rightarrow 0 \atop \eta \rightarrow \infty} \left[ \oint_{0} \frac{{\rm d}z}{2 \pi i z} \phi(z) \log(z) - \oint_{\infty} \frac{{\rm d}z}{2 \pi i z} \phi(z) \log(z) + \int_{-\eta}^{-\epsilon} \frac{{\rm d}z}{z} \phi(z) \right]\\
&= 0 \, ,
}
which is similar to the constraint in \cite{Hands:2010zp} except that there is no pole resulting from taking the sum over $n$ in the equation of motion since we have truncated it. Plugging in $\phi(z)$ from (\ref{res-mu1}) results in an expression which constrains the Polyakov lines and other observables to satisfy the $SU(N_c)$ constraint,
\EQ{
{\cal N} = \frac{(1+x) \log r}{(1-x)-(1+x) \log\left( \frac{2}{1+x} \right)} \, ,
\label{n-quark-sun}
}
which is clearly $0$ for $r=1$. It is now possible to solve for ${\cal N}$, $r$, and $x$ as a function of $T R$. Plugging $r$ from (\ref{r-eq-mu}) into (\ref{n-quark}) and (\ref{n-quark-sun}) gives two equations for ${\cal N}$ as a function of $x$, which we refer to ${\cal N}_1$, and ${\cal N}_2$, respectively. Setting ${\cal N}_1 = {\cal N}_2$ gives the equation for $x(T)$, which can be used to solve for the Polyakov lines (\ref{polys-qcd1-1}).

\begin{figure}[t]
  \hfill
  \begin{minipage}[t]{.49\textwidth}
    \begin{center}
\includegraphics[width=0.99\textwidth]{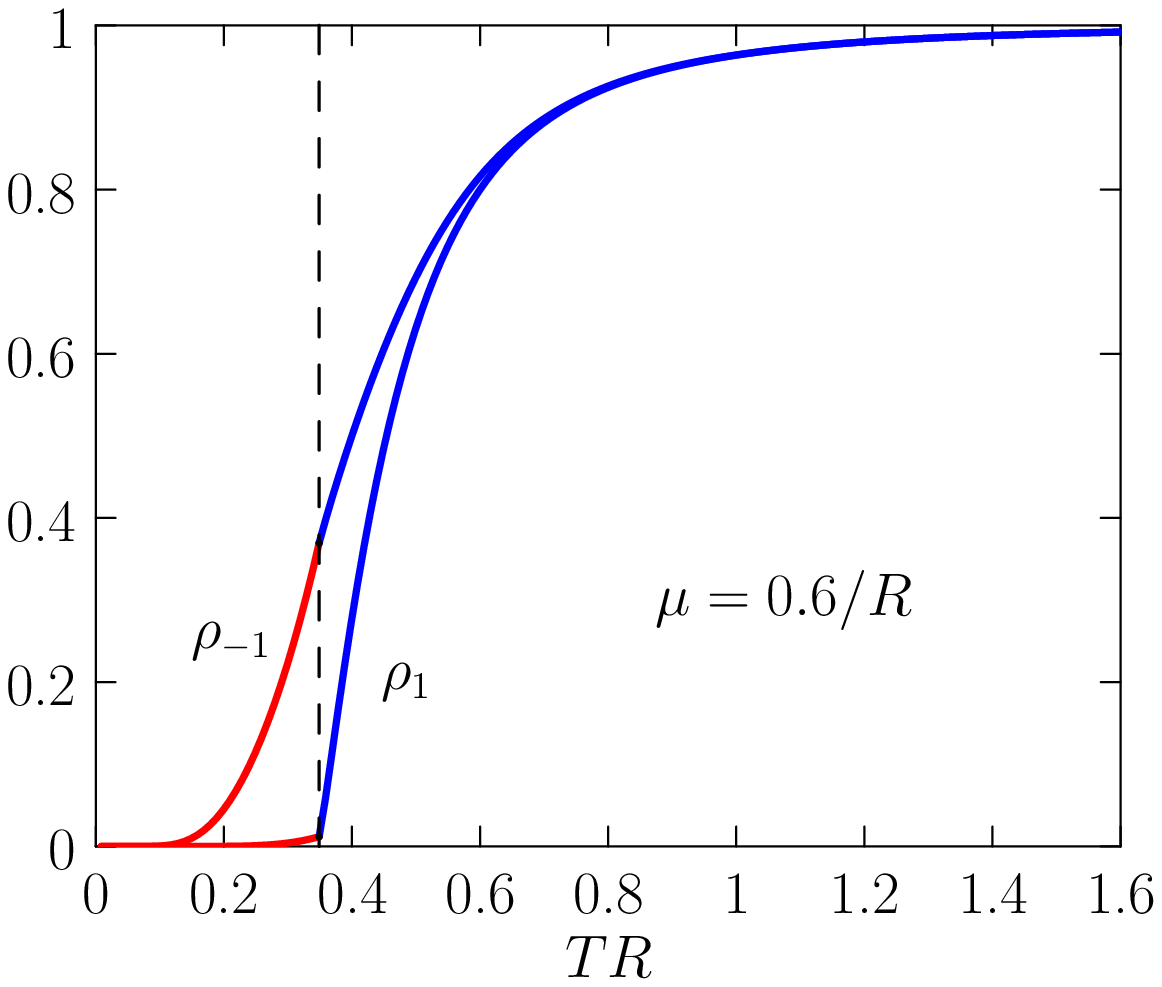}
    \end{center}
  \end{minipage}
  \hfill
  \begin{minipage}[t]{.49\textwidth}
    \begin{center}
\includegraphics[width=0.99\textwidth]{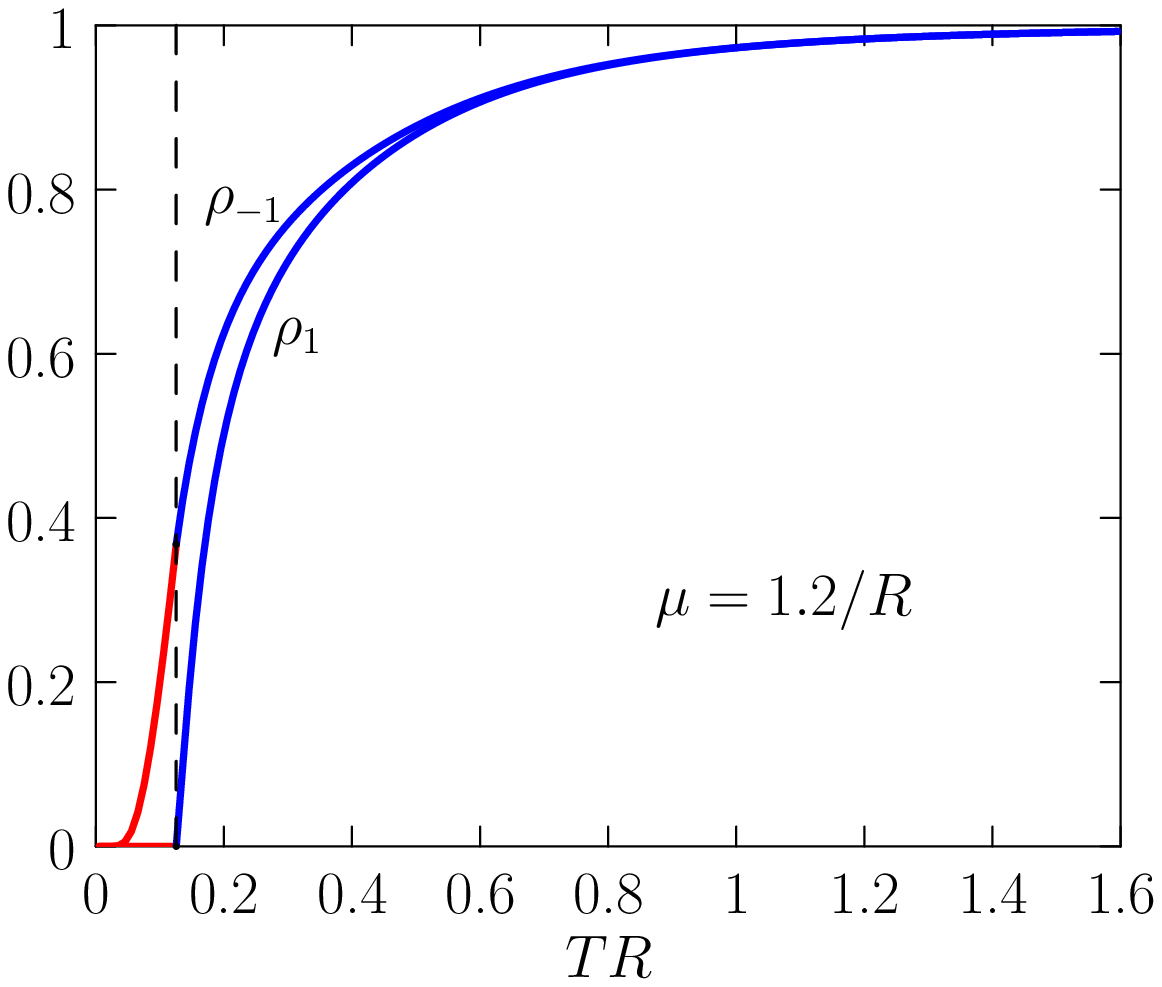}
    \end{center}
  \end{minipage}
  \hfill
\caption{Polyakov lines $\rho_1$ and $\rho_{-1}$ as a function of the temperature for $m R = 0$, $\frac{N_f}{N_c} = 1$. (Left) $\mu = 0.6/R$, (Right) $\mu = 1.2/R$.} 
\label{fig-trP1-trans}
\end{figure}

In Figure \ref{fig-trP1-trans}, the Polyakov lines $\rho_{\pm1}$ are plotted as a function of the temperature for $\mu = 0.6/R$, $\mu = 1.2/R$. The red part of the curve corresponds to the ungapped distribution, and the blue to the gapped distribution. The transition is indicated by the dotted line. The sharper transition of $\rho_1$ compared to $\rho_{-1}$ also occurred in the low temperature results in \cite{Hands:2010zp}, where the fact that $\rho_1 \ne \rho_{-1}^*$ for $\mu \ne 0$ was also observed and the extent of the difference served as an indication of the severity of the sign problem.

\begin{figure}[t]
  \hfill
  \begin{minipage}[t]{.49\textwidth}
    \begin{center}
\includegraphics[width=0.99\textwidth]{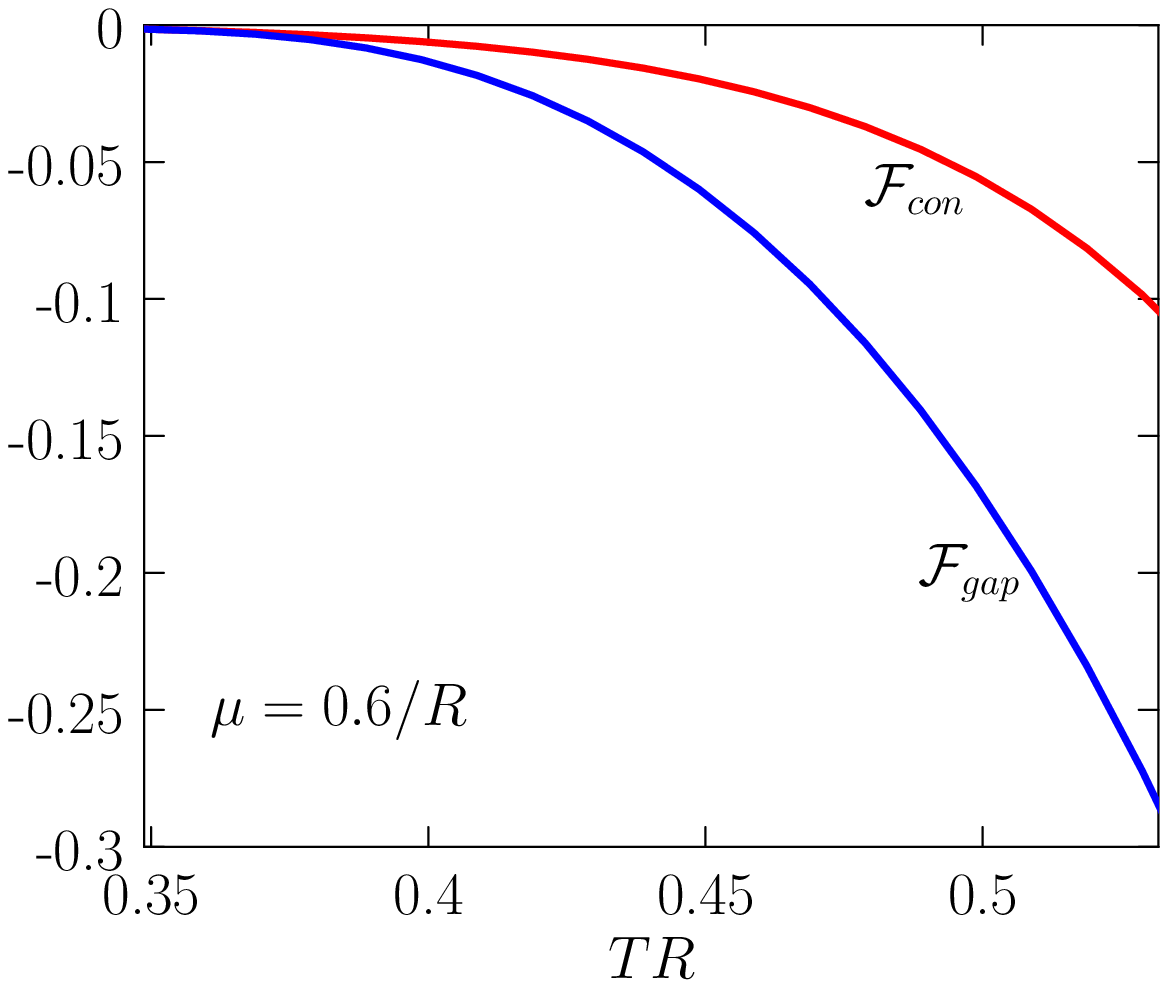}
    \end{center}
  \end{minipage}
  \hfill
  \begin{minipage}[t]{.49\textwidth}
    \begin{center}
\includegraphics[width=0.99\textwidth]{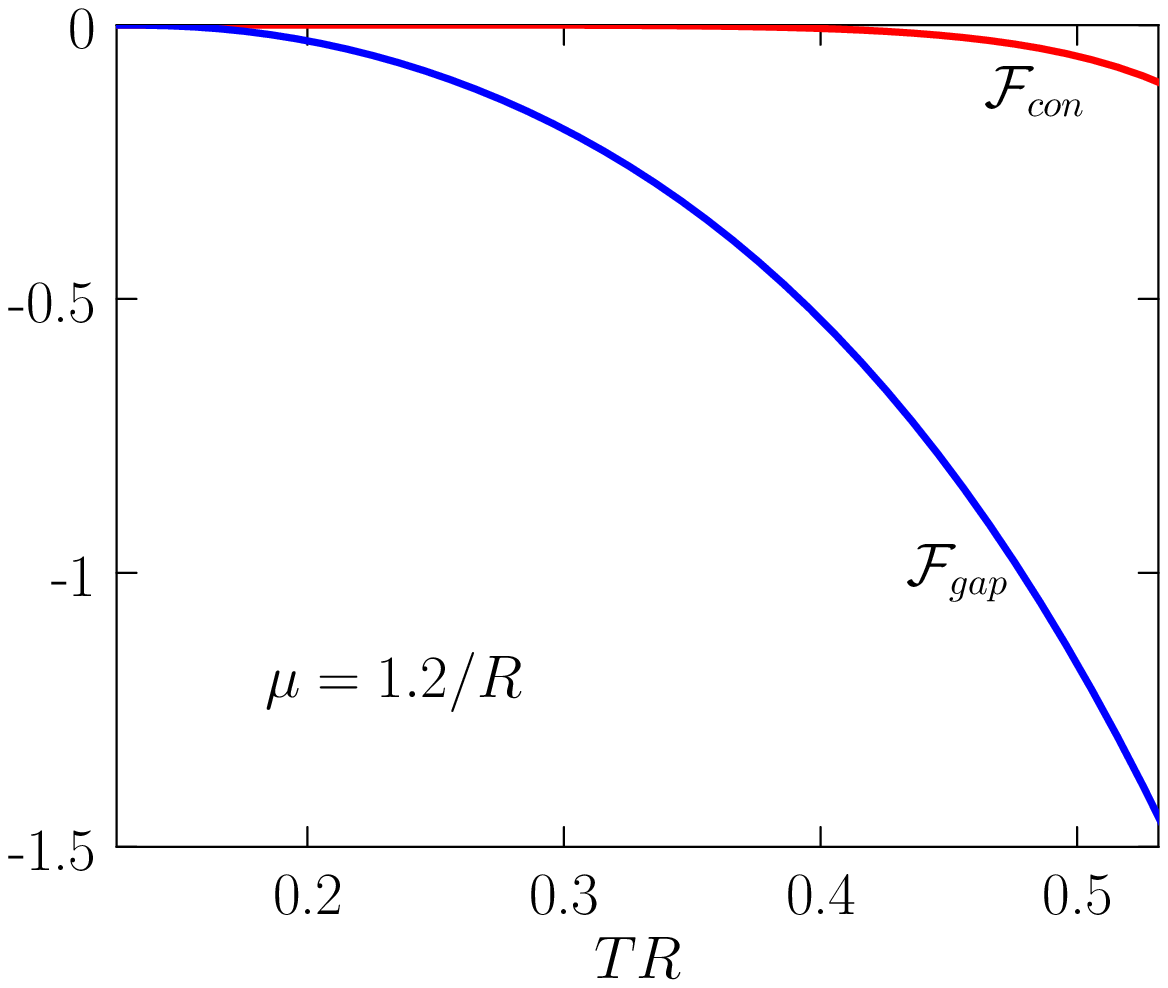}
    \end{center}
  \end{minipage}
  \hfill
\caption{Free energy of the gapped and continuous distributions for the range of temperatures $T_{c}^{(1)} < T < T_{c}^{(2)}$ in which both are possible, for $m R = 0$, $\frac{N_f}{N_c} = 1$. (Left) $\mu = 0.6/R$. (Right) $\mu = 1.2/R$.} 
\label{fig-freeE-trans}
\end{figure}

As the temperature is increased there is a certain critical temperature, $T_{c}^{(1)}$ at which it becomes possible for the distribution of Polyakov line eigenvalues to develop a gap. This occurs at the temperature at which $x(T)$ develops a solution, in the range $[-1,1]$, which is not $-1$. However, up to a second critical temperature, $T_{c}^{(2)}$, the ungapped solution is also still possible. This second critical temperature is given by the solution of (\ref{qcd0-Tc}), which is $\mu$-independent because the transition occurs from a phase in which ${\cal N} = 0$, where the Polyakov line can be shifted to absorb the $\mu$-dependence. Therefore, to determine the temperature at which the gapped solution takes over it is necessary to compare the free energies for the gapped and ungapped distributions for the range of temperatures, $T_{c}^{(1)} < T < T_{c}^{(2)}$, in which both are possible. This is done in Figure \ref{fig-freeE-trans} for $\mu = 0.6/R$ and $\mu = 1.2/R$, using
\EQ{
{\cal F} \equiv \frac{1}{N_c^2} F = \frac{1}{N_c^2} T S \simeq T \left[ (1 - {\boldsymbol z}_{v1}) \rho_1 \rho_{-1} - \frac{N_f}{N_c} {\boldsymbol z}_{f1} \left( \rho_1 e^{\mu \beta} + \rho_{-1} e^{-\mu \beta} \right) \right] \, .
}
It is clear from Figure \ref{fig-freeE-trans} that for both $\mu = 0.6/R$ and $\mu = 1.2/R$ the gapped solution results in a lower free energy for the entire range of temperatures in which both the gapped and ungapped solutions are possible. Therefore as the temperature is increased the gapped solution will take over as soon as it exists. We have compared the free energies for several other values of $\mu$ and this seems to hold for the full range $\mu = 0$ to $\mu \approx \varepsilon_{f1}$.

\begin{figure}[t]
\center
\includegraphics[width=9cm]{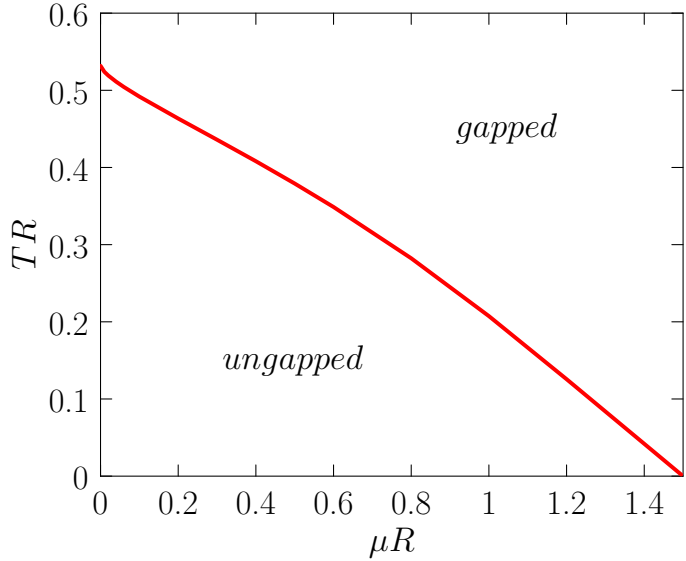}
\caption{Phase diagram based on the distribution of the eigenvalues of the Polyakov line for $m R = 0$, $\frac{N_f}{N_c} = 1$. The line of transitions corresponds to the critical temperatures at which a gap forms in the distribution. The transition at $\mu = 0$ is at least fifth order. For $\mu > 0$, $T \rightarrow 0$ it is third order. In between the quark number is smoothly connected across the transitions which implies that the transitions are at least second order.}
\label{fig-phase-diag-mu1}
\end{figure}

In Figure \ref{fig-phase-diag-mu1} we plot the phase diagram resulting from the Polyakov lines as a function of temperature for different values of the chemical potential, in the $\mu R$-$T R$ plane, for $\frac{N_f}{N_c} = 1$, and $m R = 0$. The line of transitions separates the region in which the distribution of the Polyakov line eigenvalues is ungapped from that in which it is gapped. The point at which the line of transitions intercepts the $T R$-axis is precisely given by the solution of (\ref{qcd0-Tc}), which is the result of \cite{Schnitzer:2004qt} for $\frac{N_f}{N_c} = 1$, $m R = 0$. As shown in section \ref{qcd} the transition at $\mu = 0$ is at least fifth order. The point at which the line of transitions touches the $\mu R$-axis is $\varepsilon_{f1} \big{|}_{m R = 0} = 1.5$, which is the result of \cite{Hands:2010zp}, where the transition is third order. What is perhaps surprising about the phase diagram is that the line of transitions has very little curvature. This is possible because the line of transitions is not first order. Continuity of the effective quark number ${\cal N}$ across the transitions implies that they are at least second order.

\begin{figure}
\begin{center}
\begin{tikzpicture}[scale=0.8]
\draw[->] (0,0) -- (6,-2);
\draw[->] (0,0) -- (0,5.5);
\draw[->] (0,0) -- (4,2.7);
\draw[-,very thick] (3,-1) .. controls (3,0) and (2.7,1.2)   ..  (2.3,1.5);
\draw[-,very thick,color=red] (3,-1) ..controls (3.2,-0.5) and (2.2,1) .. (6.5,5);
\draw[-,very thick,color=red] (4.7,-1.6) ..controls (4.7,-1.1) and (5.2,2) .. (7.5,4.7);
\draw[-,very thick] (4.4,-0.6) ..controls (4.4,-0.1) and (5.1,2.6) .. (7.4,5.3);
\draw[-,very thick] (3,-1) -- (4.4,-0.6);
\draw[-,very thick] (4.7,-1.6) -- (4.4,-0.6);
\draw[-,very thick] (2.3,1.5) ..controls (2.3,2.2) and (3,2) .. (3,6);
\draw[very thin,pattern=north west lines,dashed] (3,-1) -- (4.4,-0.6) ..controls (4.4,-0.1) and (5.1,2.6) .. (7.4,5.3) -- (6.5,5) .. controls (2.2,1) and (3.2,-0.5)  .. (3,-1);
\draw[very thin,pattern=north west lines,dashed] (3,-1) .. controls (3,0) and (2.7,1.2)   ..   (2.3,1.5) .. controls (2.3,2.2) and (3,2) .. (3,6) .. controls (3.8,5.9) and (6.2,5.4) .. (6.5,5) .. controls (2.2,1) and (3.2,-0.5)  .. (3,-1);
\draw[very thin,pattern=north west lines,dashed] (4.7,-1.6) ..controls (4.7,-1.1) and (5.2,2) .. (7.5,4.7) -- (7.4,5.3) ..controls (5.1,2.6) and (4.4,-0.1) .. (4.4,-0.6) -- (4.7,-1.6);
\draw[very thick,densely dashed] (4.7,-1.6) -- (6,-1.4);
\draw[very thick,densely dashed] (6.2,-1.8) -- (6,-1.4);
\draw[very thick,densely dashed] (7.5,4.7) -- (8.3,4.9);
\draw[very thick,densely dashed] (8.4,4.6) -- (8.3,4.9);
\node at (0,6) (a1) {$m$};
\node at (6.3,-2.2) (a2) {$\mu$};
\node at (4.9,2.7) (a3) {$T$};
\node at (1.3,3) (a4) {confined};
\node at (7.3,1) (a5) {deconfined};
\node at (3,-2) (a6) {confined};
\draw[->] (a6) -- (4.1,-1.1);
\draw[->] (a6) -- (5.6,-1.7);
\end{tikzpicture}
\caption{\small A qualitative picture of the phase transition in $\mu$-$T$-$m$ space at large $N$. For $\mu=0$, as $m\to\infty$ the transition goes to the critical temperature for the Yang-Mills theory. For $T=0$ the lowest transition is set by the energy of lowest fermion mode $\mu = \sqrt{m^2+9/(4R^2)}$. There are further transitions as $\mu$ increases creating a sawtooth pattern and forming lines (shown in red) in the $T=0$ plane at $\mu=\sqrt{m^2+(\ell+\tfrac12)^2/R^2}$, $\ell=1,2,\ldots$, \cite{Hands:2010zp}.}
\label{phase-3d}
\end{center}
\end{figure}
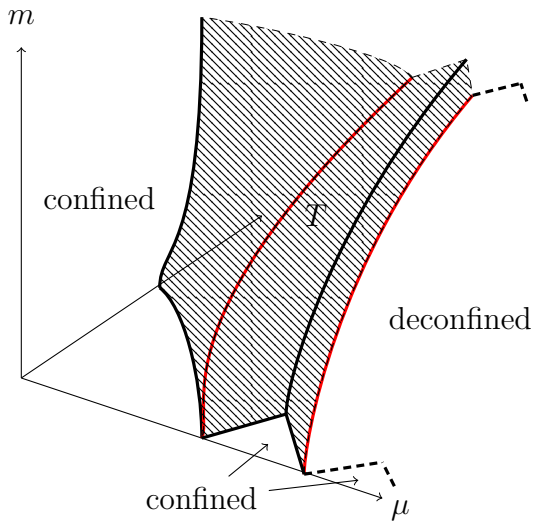

It is possible to qualitatively extend the phase diagram of Figure \ref{fig-phase-diag-mu1} into the region of nonzero quark mass. The critical behavior with large mass at zero chemical potential is known to approach the Yang-Mills theory result, and at large mass and zero temperature the critical chemical potential goes to $\mu = \varepsilon_{f1} = \sqrt{m^2+9/(4R^2)}$, as found in \cite{Hands:2010zp}. The resulting phase diagram is sketched in Figure \ref{phase-3d}.
\section{Strong coupling expansion of lattice QCD with $\mu \ne 0$}
\label{lattice-results}

Making the substitutions (\ref{change-vars}), the Polyakov lines in (\ref{polys-qcd1-1}) can be translated into their forms for the lattice strong coupling expansion,
\EQ{
\frac{1}{N_c} \langle W \rangle = \frac{h (1-x) e^{- \mu/T} \left[ 8+4 e^{2 \mu/T} r^{2} (1+x)- J D (1-x)^2 (3+x)\right]}{16 - J D (1-x) \left[16- J D (1-x)^2 (3+x) \right]} \, ,
\label{poly1-lat}
}
\EQ{
\frac{1}{N_c} \langle W^{\dagger} \rangle = \frac{h (1-x) e^{\mu/T} \left[ 8+4 e^{-2 \mu/T} r^{-2} (1+x)- J D (1-x)^2 (3+x)\right]}{16 - J D (1-x) \left[16- J D (1-x)^2 (3+x) \right]} \, ,
\label{poly2-lat}
}
where from (\ref{r-eq-mu})
\EQ{
r = \frac{4 - J D (1-x)(3+x) + \sqrt{\left[ 4 - J D (1-x)(3+x) \right]^2 - 16 (1-x)^2 h^2}}{4 (1-x) e^{\mu \beta} h} \, ,
\label{r-lat}
}
and $x$ is obtained by equating (\ref{n-quark}), in the form
\EQ{
{\cal N} = (1+x) \left[ -\frac{1}{1-x} + \frac{4 h r^{-1} e^{-\mu \beta} \left[ J D (1- x^2) +2 r^2 e^{2 \mu \beta} (2 - J D (1-x)) \right]}{16 - J D (1-x) \left[ 16 - J D (1-x)^2 (3+x) \right]} \right] \, ,
\label{n-lat}
}
with (\ref{n-quark-sun}).

\begin{figure}[t]
  \hfill
  \begin{minipage}[t]{.49\textwidth}
    \begin{center}
\includegraphics[width=0.99\textwidth]{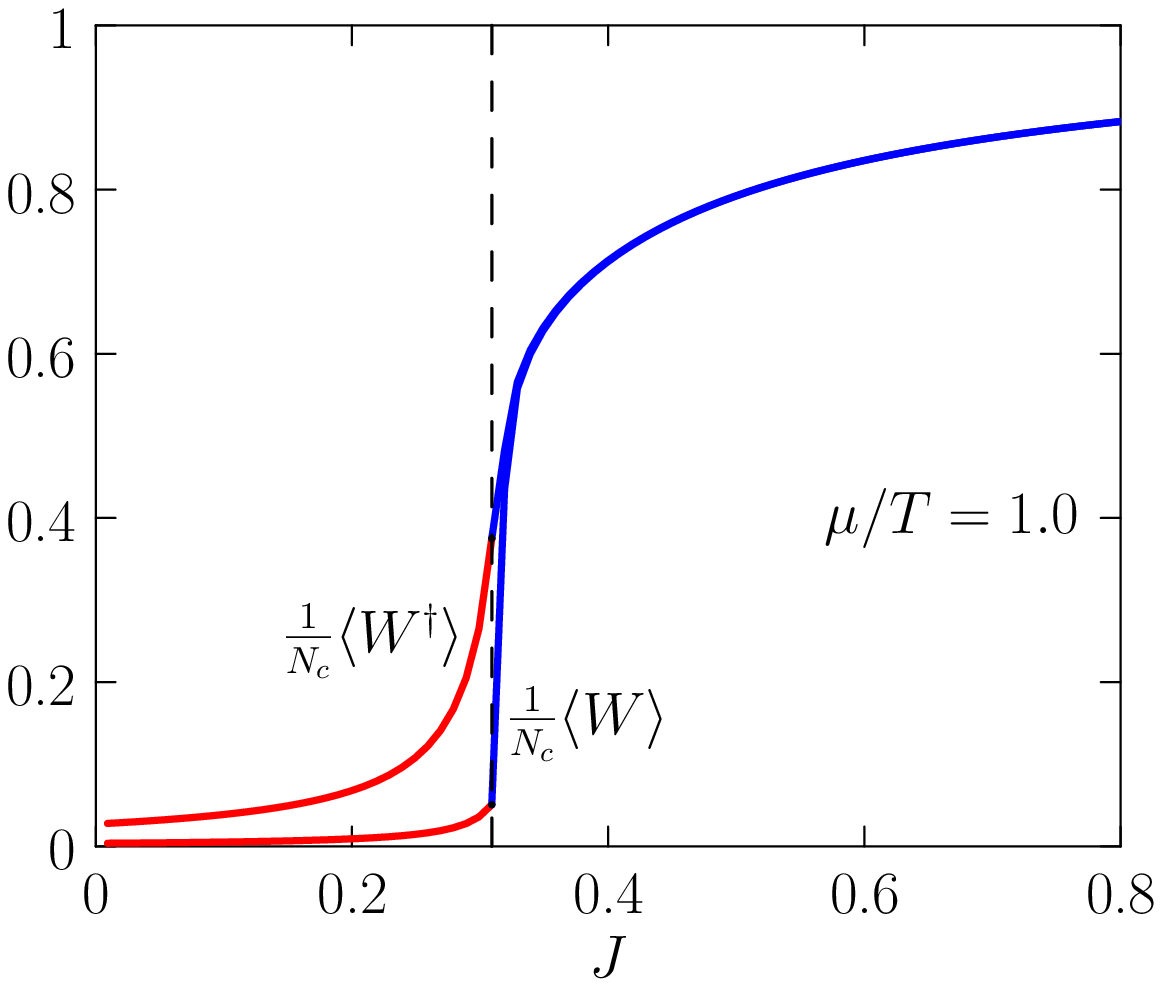}
    \end{center}
  \end{minipage}
  \hfill
  \begin{minipage}[t]{.49\textwidth}
    \begin{center}
\includegraphics[width=0.99\textwidth]{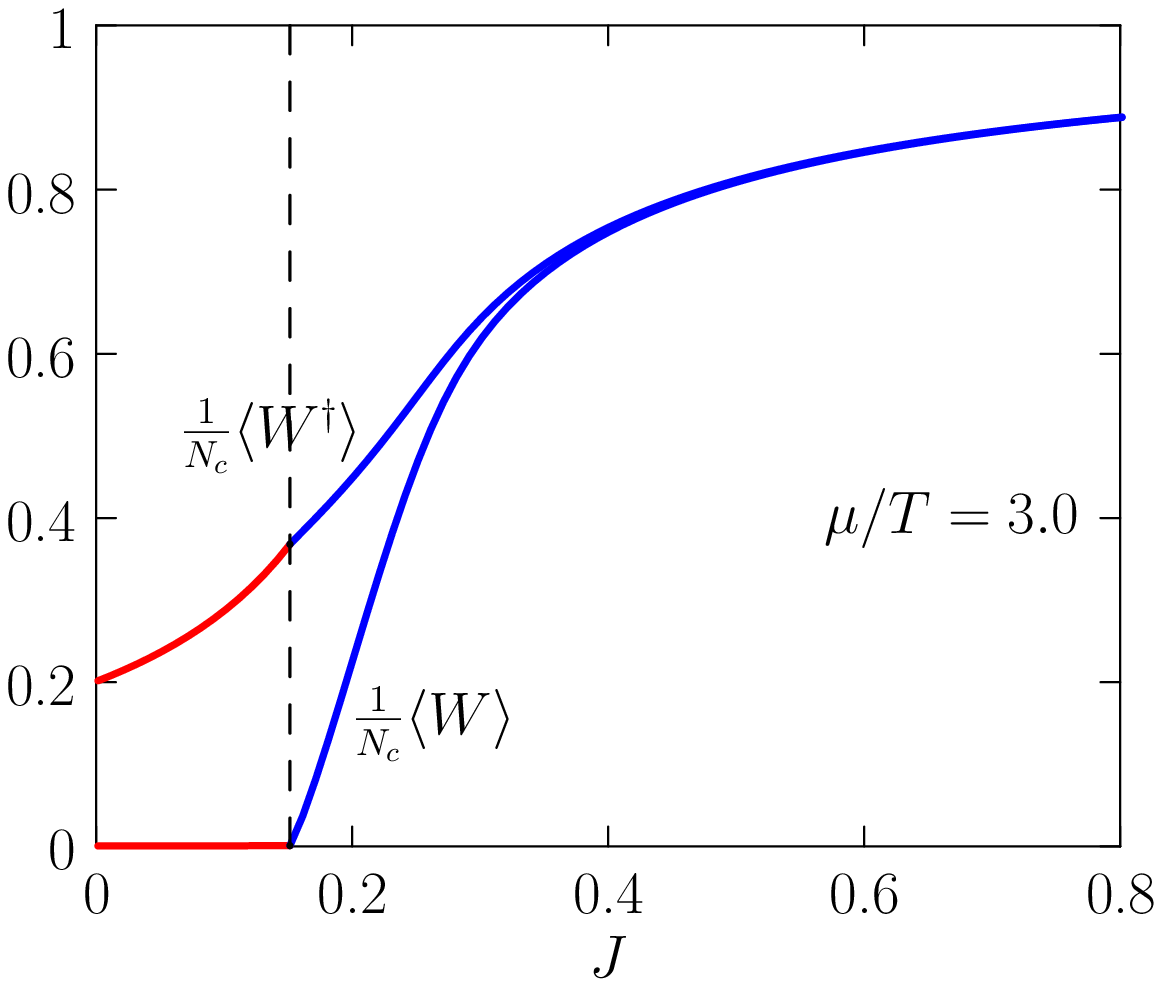}
    \end{center}
  \end{minipage}
  \hfill
\caption{Polyakov lines $\frac{1}{N_c} \langle W \rangle$ and $\frac{1}{N_c} \langle W^{\dagger} \rangle$ as a function of $J$ for $h=0.01$. (Left) $\mu/T = 1.0$, (Right) $\mu/T = 3.0$.} 
\label{fig-trP-lat}
\end{figure}

The Polyakov lines (\ref{poly1-lat}), (\ref{poly2-lat}), are plotted in Figure \ref{fig-trP-lat} for $\mu / T = 1.0$ and $\mu / T = 3.0$, as a function of $J$. We have set $h = 0.01$, which corresponds, approximately, to $m a = 0.05$, $D = 3$, $N_{\tau} = 4$, and $\frac{N_f}{N_c} = \frac{4}{3}$, which we have chosen in the hope of comparing to the simulation results in \cite{Fodor:2001au,D'Elia:2002gd,Azcoiti:2005tv,Kratochvila:2005mk} (for a review see \cite{deForcrand:2010ys}). However, in the very strong coupling limit $J = 2 \left( \frac{\beta_{lat}}{2 N_c^2} \right)^{N_{\tau}}$ \cite{Billo:1994ss} such that the transitions from the strong coupling expansion occur for coupling strengths which are too small, $(\frac{\beta_{lat}}{N_c^2})_c \simeq 1.2$ for $\mu / T \lapprox 1.5$, while simulation results indicate that the transitions occur for $(\frac{\beta_{lat}}{N_c^2})_c \simeq 0.5$. It would be worthwhile to determine if corrections to the leading contribution in the strong coupling and hopping parameter expansions result in a lower $(\frac{\beta_{lat}}{N_c^2})_c$.

\begin{figure}[t]
  \hfill
  \begin{minipage}[t]{.49\textwidth}
    \begin{center}
\includegraphics[width=0.99\textwidth]{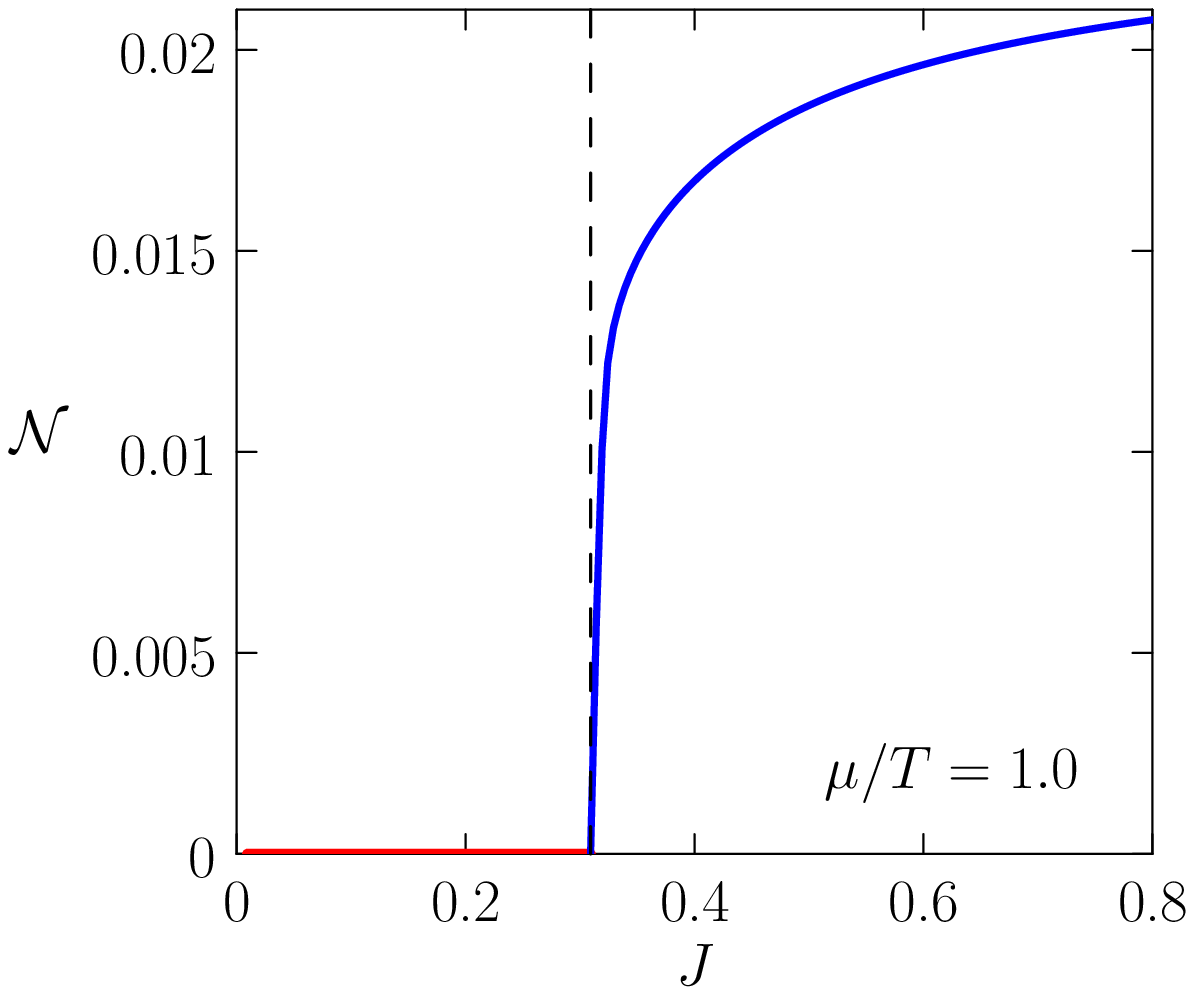}
    \end{center}
  \end{minipage}
  \hfill
  \begin{minipage}[t]{.49\textwidth}
    \begin{center}
\includegraphics[width=0.99\textwidth]{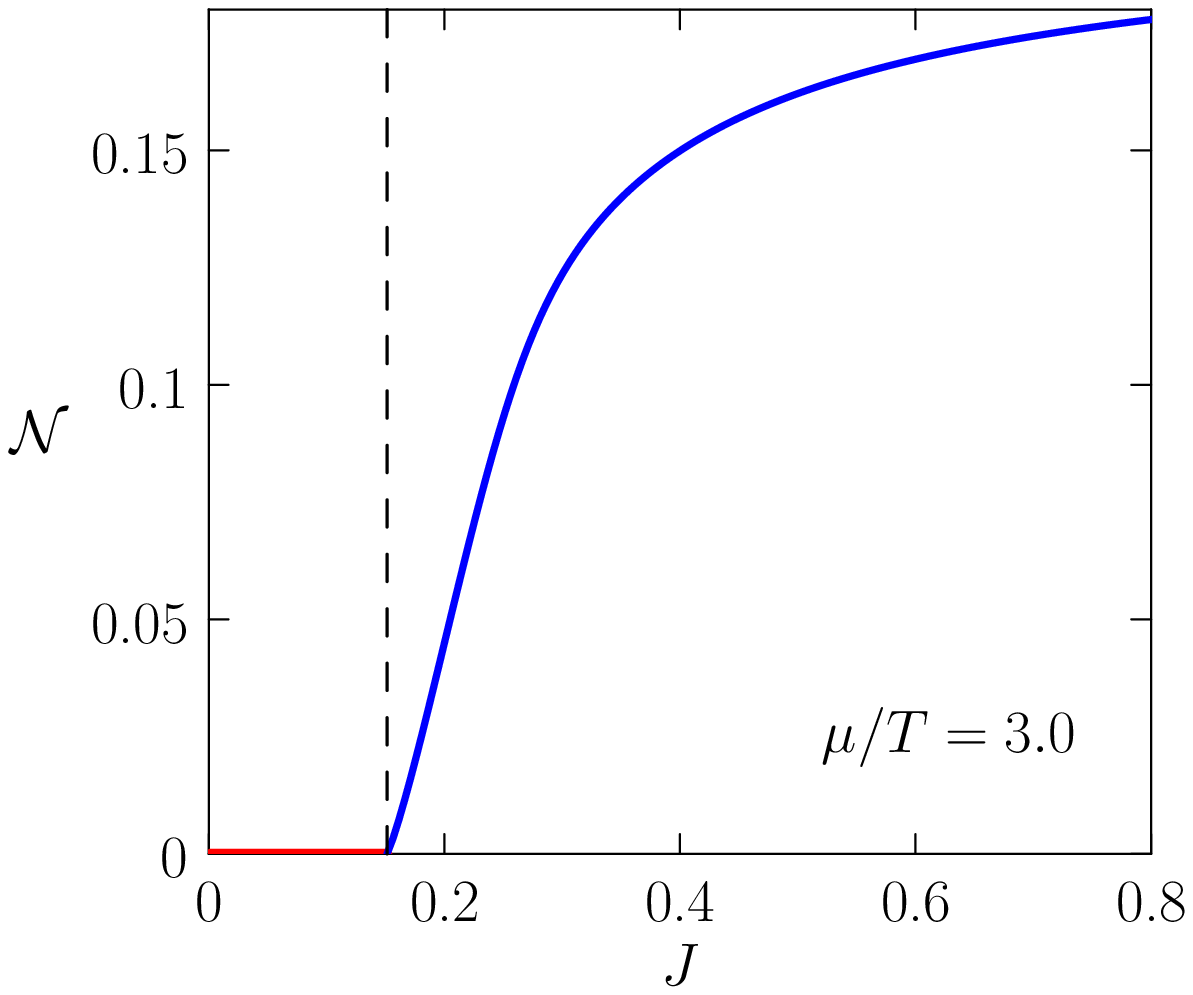}
    \end{center}
  \end{minipage}
  \hfill
\caption{Effective quark number ${\cal N} = \frac{1}{N_c^2} N_q$ for $h=0.01$. (Left) $\mu/T = 1.0$. (Right) $\mu/T = 3.0$. Note that the scale is not the same.} 
\label{fig-n-quark-lat}
\end{figure}

The effective quark number, ${\cal N} = \frac{1}{N_c^2} N_q$, obtained from (\ref{n-lat}) is plotted in Figure \ref{fig-n-quark-lat} for $\mu / T = 1.0$ and $\mu / T = 3.0$, where the red part of the curve corresponds to the ungapped distribution and the blue part to the gapped distribution. It is clear that the value of $J$ at which the distribution develops a gap corresponds to the value at which the quark number becomes non-zero. Since ${\cal N}$ is always connected during the transition from the ungapped to the gapped phase the transitions are at least second order. Note that in the lower temperature and larger $\mu$ results in \cite{Hands:2010zp} it was also found that ${\cal N} \ne 0$ in the low temperature regions of the ungapped phase when $\mu > \varepsilon_{f1}$, where the sawtooth pattern of confinement-deconfinement transitions in Figure \ref{phase-3d} takes place.

\begin{figure}[t]
\center
\includegraphics[width=9cm]{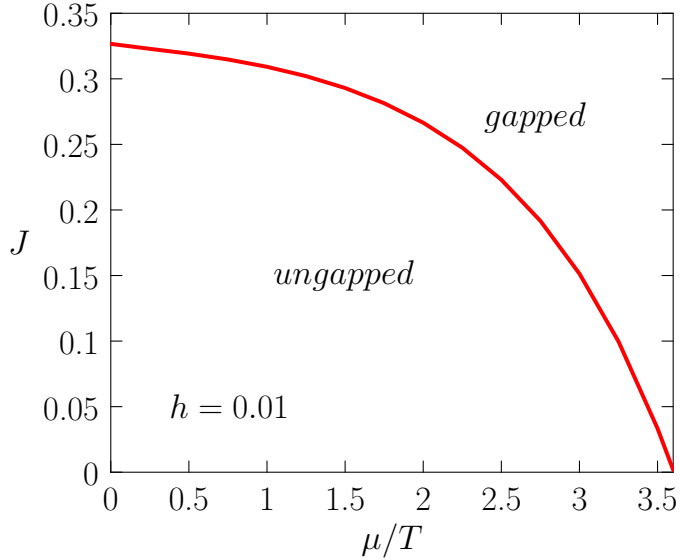}
\caption{Phase diagram of large $N_c$ QCD from the lattice strong coupling expansion for $h=0.01$. At $\mu / T = 0$ the transition is at least fifth order. For $\mu / T > 0$ the continuity of the effective quark number ${\cal N}$ implies that the line of transitions is at least second order.}
\label{fig-phase-diag-lat}
\end{figure}

The phase diagram is plotted as a function of $J$ and $\mu / T$ in Figure \ref{fig-phase-diag-lat}, for $h = 0.01$. The deconfinement line of transitions is defined by the points at which $x$ develops a solution which is not $-1$, or equivalently by the points at which ${\cal N}$ becomes nonzero. The line of transitions separates the region in which the distribution of the Polaykov line eigenvalues is ungapped, from that in which it is gapped, and it touches the $J$ axis precisely at $J_c = \frac{1}{D} (1 - 2 h) \simeq 0.32667$, which is the $\mu = 0$ result in \cite{Damgaard:1986mx}. At this point the transition is at least fifth order as shown in section \ref{qcd}. For $\mu / T > 0$ continuity of the quark number across the transitions implies that they are at least second order. The point at which the line of transitions intersects the $\mu / T$ axis is given by $(\mu / T)_c \simeq 3.6$.

\begin{figure}
\begin{center}
\begin{tikzpicture}[scale=0.6]
\draw[->] (0,0) -- (-6,-6);
\draw[->] (0,0) -- (0,5);
\draw[->] (0,0) -- (5,0); 
\draw[-,very thick] (0,4) .. controls (-0.2,3.4) and (-3,-2) .. (-5.5,-5); 
\draw[-,very thick] (4,0) -- (-2,-6); 
\draw[-,very thick] (3.7,0.3) .. controls (3,-0.6) and (-3,-3.7) .. (-4.7,-4); 
\draw[-,very thick] (3.4,0.6) .. controls (2.8,-0.5) and (-3,-2.4) .. (-3.55,-2.3); 
\filldraw[black] (3.4,0.6) circle (4pt);
\filldraw[black] (-3.55,-2.3) circle (4pt);
\draw[-,very thick] (2,2) .. controls (1.5,1.5) and (-1.2,0.6) .. (-1.75,0.6); 
\draw[-,very thick] (1,3) .. controls (0.6,2.7) and (-0.7,2.4) .. (-0.75,2.5); 
\draw[-,very thick] (4,0) -- (0,4); 
\node at (0,5.5) (a1) {$h$};
\node at (5.5,0) (a2) {$J$};
\node at (-6.5,-6.5) (a3) {$\mu/T$};
\node at (5,3.5) (a4) {$J=\frac1D(1-2h)$};
\draw[->] (a4) -- (2.4,1.7);
\node at (7.5,2) (a5) {$(h=0.01,J=0.3267)$};
\draw[->] (a5) -- (3.65,0.7);
\node at (-2.5,4) (a6) {$h=\frac12$};
\draw[->] (a6) -- (-0.2,4);
\node at (-9,-2) (a7) {$(h=0.01,\mu/T=3.6)$};
\draw[->] (a7) -- (-3.75,-2.22);
\node at (4.5,-4) (a8) {$J=1/D$};
\draw[->] (a8) -- (2.3,-2);
\node at (-9,-5) (a9) {asymptotes to 0};
\draw[->] (a9) -- (-5.6,-4.9);
\end{tikzpicture}
\caption{\small A qualitative picture of the phase transitions in $\mu/T$-$J$-$h$ space. For $\mu/T=0$, the critical line is given by $J = \frac{1}{D}(1-2h)$ \cite{Damgaard:1986mx}, with $h_{max} = \frac{1}{2}$. For $h = 0$ the transition is the Yang-Mills result $J = \frac{1}{D}$. For $J=0$ the transition asymptotes to $h = 0$ as $\mu/T \rightarrow \infty$. The slice of transitions at $h = 0.01$ as a function of $J$ and $\mu / T$ is plotted in Figure \ref{fig-phase-diag-lat}.}
\label{lat-phase-3d}
\end{center}
\end{figure}
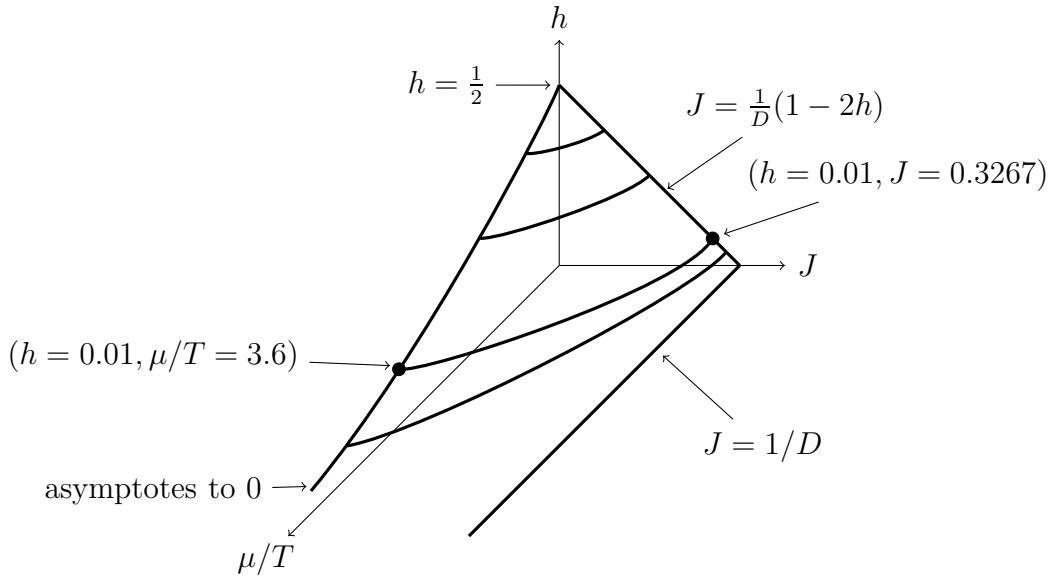

Since it is difficult to compare quantitatively with lattice simulation results, even when using the same value of $h$, it is helpful to consider what happens qualitatively as $h$ is varied along with $J$ and $\mu / T$, which is sketched in Figure \ref{lat-phase-3d}. At $\mu / T = 0$ the critical behavior is that of \cite{Damgaard:1986mx}, $J = \frac{1}{D} \left( 1 - 2 h \right)$, therefore the maximum value of $h$ is $\frac{1}{2}$, at which the line of transitions as a function of $J$ and $\mu/T$ reduces to a point. For $h = 0$, or $m a = \infty$, the critical behavior reduces to the Yang-Mills result $J = \frac{1}{D}$. Therefore as $h$ is reduced, or as $m a$ is increased, the line of transitions at zero $J$ is stretched along the $\mu / T$ axis, such that the Yang-Mills result is approached in the $h \rightarrow 0$ limit.

\section{Discussion}

We have calculated the phase diagram of large $N_c$ QCD with small chemical potential, $\mu \lapprox \varepsilon_{f1}$, and moderate temperature from one-loop perturbation theory on $S^1 \times S^3$, and to leading order in the lattice strong coupling and hopping parameter expansions, by solving a single matrix model. It would be interesting to investigate whether it is possible to extend both the weak and strong coupling results to the next order, or beyond, using a single matrix model and a suitable change of parameters, by including also terms with multiply wound Polyakov lines, $\rho_{\pm n}$, for $n > 1$, which are also known to be relevant at higher orders in the strong coupling expansion. Along these lines, one could see how the phase diagram is extended into the region of more moderate quark masses by including contributions from ${\boldsymbol z}_{v2}$, ${\boldsymbol z}_{v3}$, etc., on $S^1 \times S^3$, or how the phase diagram is extended into the region of larger chemical potentials by including the contributions from ${\boldsymbol z}_{f2}$, ${\boldsymbol z}_{f3}$, etc.. For the lattice theory the former would be equivalent to considering additional terms in the strong coupling expansion, and the latter would be equivalent to including more terms in the hopping parameter expansion (following, for example, \cite{Langelage:2010yn,Fromm:2011qi}). At low temperatures one could expect to reproduce the sawtooth pattern of confinement-deconfinement transitions found in \cite{Hands:2010zp}.

In this work we have considered only the ungapped and single gapped phases where the gap opens on the negative real-axis, as expected for the ordinary deconfined phase. It might also be interesting to consider multi-gap distributions, which were considered in \cite{Jurkiewicz:1982iz} for Yang-Mills theory.

It might be that approximations exist which lead to a correspondence of matrix models at weak and strong coupling for other theories. The qualitative similarity of weak-coupling results from $S^1 \times S^3$ for QCD with adjoint fermions in \cite{Hollowood:2009sy}, and lattice simulation results in \cite{Cossu:2009sq}, seems to suggest this possibility. However, in this case it could be necessary to consider higher order terms from the strong coupling and/or hopping parameter expansion to match the phase diagrams since multi-gap solutions are possible, which are a result of transitions in terms with ${\boldsymbol z}_{vn}$ and ${\boldsymbol z}_{fn}$ for $n > 1$ in the case of the weakly-coupled theory on $S^1 \times S^3$.

The behavior of the order of transitions as a function of the chemical potential for ordinary QCD with $N_c = 3$ is a currently debated issue. If a critical endpoint exists at some finite chemical potential where the line of transitions would become first order then its location should be bound from below by the radius of convergence of a Taylor expansion around $\mu = 0$ of the pressure, the first few terms of which can be obtained from lattice simulations, where the radius of convergence is related to the quark number susceptibility \cite{Gavai:2004sd}. However, the predicted location of the lower bound on a possible critical endpoint is thought to lie out of the current reach of both lattice simulations and experiment \cite{Gupta:2009mu}. It is interesting that there are simulation results from QCD, and the $3$-state Potts model, performed in the limit of small chemical potentials, which seem to predict that the transition becomes weaker with increasing $\mu$, suggesting that a critical endpoint might not be observed \cite{deForcrand:2007rq,Kim:2005ck}. For these reasons it would be interesting to consider further corrections from ${\boldsymbol z}_{vn}$, ${\boldsymbol z}_{fn} \ne 0$ for $n = 2, 3, ...$ to more precisely obtain the order of the line of transitions as a function of $\mu$. In Section \ref{ym-corrections} we found that these corrections are only barely perceptible in $\rho_1$ for Yang-Mills theory, and that the location and order of the first order phase transition remains unchanged. However, it remains possible, and should be checked, if such small corrections to $\rho_1$ are able to perturb the $N_f \ne 0$ system enough to change the order of the phase transition.

\section{Acknowledgements}

We would like to thank Philippe de Forcrand, Simon Hands, Jens Langelage, Tiago Nunes da Silva, Marco Panero, Andr{\'a}s Patk{\'o}s, Owe Philipsen, Kim Splittorff, and Wolfgang Unger for discussions and Jan Rosseel for looking over the manuscript.




\end{document}